\documentclass[journal]{IEEEtran}
\usepackage{mathtools}  
\usepackage{amssymb}  
\usepackage{bm}  
\usepackage{outlines}  
\usepackage{bbm}
\usepackage[mathscr]{euscript}
\usepackage{xcolor}
\usepackage[bookmarks=false,hyperfigures=false,hidelinks]{hyperref}
\usepackage{url,doi}
\usepackage{booktabs}  
\usepackage{subcaption}  
\usepackage{pgfplots}  
\usepackage{tikz}  
\usepackage{acronym}
\usepackage[ruled]{algorithm2e}
\usepackage[utf8]{inputenc}
\usepackage[export]{adjustbox}

\acrodef{ADMM}[ADMM]{alternating direction method of multipliers}
\acrodef{DRS}[DRS]{Douglas-Rachford splitting}
\acrodef{SRB}[SRB]{spectral radius bound}

\renewcommand{\vec}[1]{\bm{#1}}

\DeclareMathOperator*{\argmin}{argmin}

\newcommand{\ip}[2]{\langle #1, #2 \rangle}

\newcommand{\x}{{\vec{x}}}  
\newcommand{\z}{{\vec{z}}}  
\renewcommand{\d}{\vec{d}}
\newcommand{\y}{{\vec{y}}}  

\newcommand{\A}{{\vec{A}}}  
\newcommand{\B}{{\vec{B}}}  
\newcommand{\D}{{\vec{D}}}  
\renewcommand{\c}{{\vec{c}}}  

\newcommand{\Q}{{\vec{Q}}}  
\newcommand{\R}{{\vec{R}}}  
\newcommand{\q}{{\vec{q}}}  
\renewcommand{\r}{{\vec{r}}}  

\renewcommand{\c}{{\vec{c}}}  

\newcommand{\afH}{{\vec{H}_{\boldsymbol{\rho}}}}  
\newcommand{\afh}{{\vec{h}_{\boldsymbol{\rho}}}}  

\newcommand{\FF}{{\mathbb{F}}}  
\newcommand{\GG}{{\mathbb{G}}}  

\newcommand{\eps}{\vec{\epsilon}}  

\renewcommand{\v}{{\vec{v}}}  

\newcommand{\w}{{\vec{w}}}  

\DeclarePairedDelimiterX{\normsz}[1]{\lVert}{\rVert}{#1}
\DeclarePairedDelimiterX{\parensz}[1]{(}{)}{#1}

\newtheorem{definition}{Definition}[section]
\usepackage{graphicx}

\begin{document}
\title{An Adaptive Multiparameter Penalty Selection Method for Multiconstraint and Multiblock ADMM}
\author{{L}uke Lozenski, \IEEEmembership{Graduate Student Member, IEEE}, Michael T. McCann, \IEEEmembership{Member, IEEE}  and \\ Brendt Wohlberg, \IEEEmembership{Fellow, IEEE}
\thanks{Luke Lozenski is with the Oden Institute for Computational Engineering and Sciences,  The University of Texas at Austin, Austin, TX 78712, USA and the Department of Electrical and Systems Engineering, Washington University in St. Louis, St. Louis, MO 63130, USA. This work began when L. Lozenski was an  intern with the Theoretical Division, Los Alamos National Laboratory, Los Alamos, NM 87545 }
\thanks{Michael McCann is with the Theoretical Division, Los Alamos National Laboratory, Los Alamos, NM 87545}
\thanks{Brendt Wohlberg is with the Computer, Computational, and Statistical Sciences Division, Los Alamos National Laboratory, Los Alamos, NM 87545}
\thanks{Further author information: (Send correspondence to  Michael T. McCann.) \\ Email: \texttt{mccann@lanl.gov}.}
\thanks{This research was supported by the Center for Nonlinear Studies, and the Laboratory Directed Research and Development program of Los Alamos National Laboratory under project number 20230771DI. LL acknowledges support from the Imaging Science Pathway fellowship funded by the NIH under grant T32 EB014855.}}

\maketitle

\begin{abstract} This work presents a new method for online selection of multiple penalty parameters for the alternating direction method of multipliers (ADMM) algorithm applied to optimization problems with multiple constraints or functionals with block matrix components. ADMM is widely used for solving constrained optimization problems in a variety of fields, including signal and image processing. Implementations of ADMM often utilize a single hyperparameter, referred to as the penalty parameter, which needs to be tuned to control the rate of convergence. However, in problems with multiple constraints, ADMM may demonstrate slow convergence regardless of penalty parameter selection due to scale differences between constraints. Accounting for scale differences between constraints to improve convergence in these cases requires introducing a penalty parameter for each constraint. The proposed method is able to adaptively account for differences in scale between constraints,
providing robustness with respect to problem transformations and initial selection of penalty parameters. It is also simple to understand and implement. 
Our numerical experiments demonstrate that the proposed method performs favorably compared to a variety of existing penalty parameter selection methods.
\end{abstract}


\section{Introduction}
\label{sec:intro}

The \ac{ADMM} is a proximal splitting algorithm \cite{combettes2011proximal}  for solving constrained optimization problems~\cite{glowinski-1975-approximation, gabay-1976-dual}. This work focuses on an \ac{ADMM} variant for solving optimization problems with multiple constraints, of the form
\begin{equation} \label{eq:multiconstraint_admm_problem}
\begin{array}{ccc}
\displaystyle{\argmin_{\x, \z}}  & f(\x) + g(\z)&  \\ 
\text{s.t.} &  \displaystyle{\A_j\x + \B_j\z = \c_j} & j = 1,\hdots, J ,
\end{array}
\end{equation} with variables
$\x \in \mathbb{R}^M$,
$\z \in \mathbb{R}^{N}$, 
constraint vectors
$\c_j \in \mathbb{R}^{P_j}$,
constraint matrices
$\A_j \in \mathbb{R}^{P_j \times M}$,
$\B_j \in \mathbb{R}^{P_j \times N}$, and
convex objective functions
$f: \mathbb{R}^M \to \mathbb{R}$, 
$g: \mathbb{R}^{N} \to \mathbb{R}$. 

The multiparameter \ac{ADMM} iterates for multiple constraints are expressed
\begin{equation}\label{eq:Multiconstraint_ADMM_x}
\begin{array}{ll}
     \x^{(k\!+\!1)} & = \displaystyle{\argmin_{\x}} \ f(\x)\\ 
     &  \displaystyle{\!+\! \sum_{j=1}^J \frac{\rho_j}{2} \left\| \A_j\x \!+\! \B_j\z^{(k)} \!-\! \c_j + \frac{\y^{(k)}_j}{\rho_j} \right\|^2}
\end{array}
\end{equation}
\begin{equation}\label{eq:Multiconstraint_ADMM_zj}
\begin{array}{ll}
     \z^{(k\!+\!1)} & = \displaystyle{\argmin_{\z}} \ g(\z)\\ 
     &  \!+\! \displaystyle{\sum_{j=1}^J \frac{\rho_j}{2} \left\| \A_j\x^{(k+1)} \!+\! \B_j\z^{} \!-\! \c_j + \frac{\y^{(k)}_j}{\rho_j} \right\|^2}
\end{array}
\end{equation}
\begin{equation}\label{eq:Multiconstraint_ADMM_yj}
    \begin{array}{ll}
          \y^{(k\!+\!1)}_j & = \y^{(k)}_j + \rho_j \big( \A_j\x^{(k\!+\!1)} + \B_j\z^{(k\!+\!1)} - \c_j \big) ,
    \end{array}
\end{equation}  where $\{\rho_j\}_{j=1}^J \subset \mathbb{R}$ is a set of positive scalars \emph{penalty parameters}, $\| \cdot \|$ denotes the $\ell_2$ norm, and $\y_j\in \mathbb{R}^{P_j}$ is known as the $j$-th \emph{dual variable} or the $j$-th \emph{Lagrange multiplier} associated with the $j$-th constraint in~\eqref{eq:multiconstraint_admm_problem}.  To simplify notation, we let $ \argmin f$ denote a single minimizer of $f$, even when $f$ does not have a unique minimizer. 

Note that grouping all constraints into a single constraint, by vertical matrix concatenation, and utilizing a single penalty parameter recovers standard ADMM, which is known to converge under a wide variety of conditions  \cite[\S 3.2]{boyd_distributed_2011}, \cite{deng_global_2015, giselsson-2016-linear, wang-2019-global} \footnote{While this formulation uses the Euclidean $\ell_2$ norm, \ac{ADMM} can  be generalized to a wide variety of Hilbert spaces~\cite{boct2019admm, lozenski2021consensus, borgens2021admm}.}.
The iteration \eqref{eq:Multiconstraint_ADMM_x}-\eqref{eq:Multiconstraint_ADMM_yj} can be shown to
converge to a solution of \eqref{eq:multiconstraint_admm_problem} based on the equivalence of multiparameter ADMM and standard ADMM outlined in Appendix \ref{app:multi_equivalence}. 

For ease of notation, the dual variable can be expressed in a stacked vector form 
$$\y^{(k+1)} = \y^{(k)} + \D_{\boldsymbol{\rho}}( \A \x^{(k+1)} + \B \z^{(k+1)} - \c),$$ where $\y^{(k)} = \begin{pmatrix}
    (\y_1 ^{(k)})^T & \hdots & (\y_J^{(k)})^T
\end{pmatrix}^T \in \mathbb{R}^P$ is the vectorized stack of multipliers. $$\B = \begin{pmatrix}
    \B_1 \\ \vdots \\ \B_J
\end{pmatrix}, \quad \A = \begin{pmatrix}
    \A_1 \\ \vdots \\ \A_J
\end{pmatrix}, \quad \text{and} \quad \c = \begin{pmatrix}
    \c_1 \\ \vdots \\ \c_J
\end{pmatrix} $$ are the grouped constraint matrices, $\boldsymbol{\rho} = \begin{pmatrix}
    \rho_1 & \hdots & \rho_J
\end{pmatrix}^T \in \mathbb{R}^J$ is the vectorized stack of penalty parameters, and \begin{equation}\label{eq:diagonal_matrix_operator}
\begin{array}{c}
      \D_{\boldsymbol{\rho}} = \D( \boldsymbol{\rho}) = \D(\rho_1,\hdots,\rho_J)\\ 
     = \operatorname{diag}(\underbrace{\rho_1}_{P_1\text{-times}}, \hdots , \underbrace{\rho_j}_{P_j\text{-times}}, \hdots, \underbrace{\rho_J}_{P_J\text{-times}}) \in \mathbb{R}^{P\times P}
\end{array}    
\end{equation}
 is a diagonal matrix operator.

A very important subclass of multiconstraint optimization problems, as in \eqref{eq:multiconstraint_admm_problem}, is multiblock optimization problems, which involve a separable objective of several variables with one of these variables being a consensus variable. This multiblock problem then has several constraints with each constraint involving the consensus variable and one other variable\footnote{Note that this definition is distinct from those of  \cite{deng2017parallel, lin2015sublinear}, which refer to any ADMM implementation with more than two variables as ``multiblock".}. A multiblock problem is formulated as 
\begin{equation} \label{eq:multiblock_admm_problem}
\begin{array}{ccc}
\displaystyle{\argmin_{\x, \z_{1},\hdots, \z_J}}  & f(\x) + \displaystyle{\sum_{j=1}^J g_j(\z_j)}&  \\ 
\text{s.t.} &  \A_j\x + \tilde{\B_j}\z_j = \c_j & j = 1,\hdots, J ,
\end{array}
\end{equation} with variables
$\x \in \mathbb{R}^M$,
$\z_j \in \mathbb{R}^{N_j}$;
vector
$\c_j \in \mathbb{R}^{P_j}$;
matrices
$\A_j \in \mathbb{R}^{P_j \times M}$ and
$\tilde{\B_j} \in \mathbb{R}^{P_j \times N_j}$;
convex functionals
$f: \mathbb{R}^M \to \mathbb{R}$
and
$g_j: \mathbb{R}^{N_j} \to \mathbb{R}$.

Defining $\z \in \mathbb{R}^N$ for $N = \sum_{j=1}^J N_j$ as $$\z = \begin{pmatrix}
    \z_1 \\ \vdots \\  \z_J
\end{pmatrix}, \ g(\z) = \sum_{j=1}^J g_j(\z_j),$$ and $\B_j\in \mathbb{R}^{P_j \times N} $ is  the matrix such that $\B_j \z = \tilde{\B_j} \z_j$, then the multiblock form in \eqref{eq:multiblock_admm_problem} reduces to a form identical to one in \eqref{eq:multiconstraint_admm_problem}. These multiblock optimization problems naturally emerge in several computational imaging applications, such as cases with multiple regularization and data fidelity terms \cite{ramani2011splitting, akiyama2019first} or applying separation of variables for proximal based optimization \cite{wahlberg2012admm, venkatakrishnan2013plug}.  

In the multiblock case, the \ac{ADMM} constraints defined in \eqref{eq:Multiblock_ADMM_x}-\eqref{eq:Multiblock_ADMM_yj} can equivalently be defined as \begin{equation}\label{eq:Multiblock_ADMM_x}
    \begin{array}{ll}
         \x^{(k\!+\!1)} & = \displaystyle{\argmin_{\x}} \ f(\x)\\ 
         &  \!+\! \displaystyle{\sum_{j=1}^J \frac{\rho_j}{2} \left\| \A_j\x \!+\! \tilde{\B_j}\z_j^{(k)} \!-\! \c_j + \frac{\y^{(k)}_j}{\rho_j} \right\|^2 }
    \end{array}
\end{equation}
\begin{equation}\label{eq:Multiblock_ADMM_zj}
    \begin{array}{ll}
         \z_j^{(k\!+\!1)} & = \displaystyle{\argmin_{\z_j} \ g_j(\z_j)}\\ 
         &     \!+\!  \frac{\rho_j}{2} \left\| \A_j\x^{(k+1)} \!+\! \tilde{\B_j}\z_j^{} \!-\! \c_j + \frac{\y^{(k)}_j}{\rho_j} \right\|^2 
    \end{array}
\end{equation}
\begin{equation}\label{eq:Multiblock_ADMM_yj}
    \begin{array}{ll}
    \y^{(k\!+\!1)}_j & = \y^{(k)}_j + \rho_j \big( \A_j\x^{(k\!+\!1)} + \tilde{\B_j}\z_j^{(k\!+\!1)} - \c_j \big).
    \end{array}
\end{equation}  An advantage of this multiblock formulation of \ac{ADMM} is that the updates of each $\z_j$ variable can be performed independently of each other. 

A further subset of multiblock ADMM of note is consensus ADMM in which $\c_j = \vec{0}$ and $\B_j = -I$, and often $\A_j = I$,   for all $j$. Consensus ADMM can be applied for distributed and asynchronous optimization of large-scale functions \cite{zhang2014asynchronous, makhdoumi2017convergence, chang2016asynchronous, mota2013d, chang2014multi, shi2014linear}.

In practice, the rate of convergence of \ac{ADMM} algorithms is strongly dependent on the choice of penalty parameters \cite{nishihara_general_2015}. 
Moreover, whereas the standard \ac{ADMM} algorithm requires selection of a single penalty parameter,
many problems with multiple constraints, such as multiblock problems in Eq. \eqref{eq:multiblock_admm_problem}, will require some form of preconditioning applied to the constraint for fast convergence~\cite{giselsson-2014-diagonal}. 
As demonstrated in Appendix \ref{app:multi_equivalence}, utilizing a distinct penalty parameter for each constraint is equivalent to applying diagonal preconditioning, and can similarly accelerate convergence. 

In general, there are no analytic methods for determining the optimal selection of penalty parameters.  Furthermore, brute-force methods that run the \ac{ADMM} with multiple penalty parameters are computationally expensive, 
scaling exponentially in the number of penalty parameters,
and are therefore impractical for large-scale or time-sensitive problems.

Alternatively, an \ac{ADMM} implementation can utilize an adaptive penalty parameter selection criterion. In an adaptive \ac{ADMM} implementation, the penalty parameter $\rho$ or penalty parameters $\{\rho_j\}_{j=1}^J$ are replaced with iteration dependent versions, $\rho^{(k)}$ or $\{\rho_j^{(k)}\}_{j=1}^J$, via a chosen selection criterion. A number of previous works have explored selection criterion for the single parameter cases 
\cite{he-2000-alternating, wohlberg_admm_2017, xu_adaptive_2017, lorenz_non_2019, mccann_robust_2024}. This work proposes an extension of the single parameter criterion proposed by \cite{mccann_robust_2024} to an adaptive multiparameter rule. To our knowledge, the only other proposed multiparameter rule is a generalization of the BBS method \cite{xu_adaptive_2017}, discussed in \cite{xu_alternating_2019}. Adapting notation from \cite{mccann_robust_2024}, an adaptive criterion for multiple penalty parameters can be defined as a function \begin{equation} \label{eq:adaptive_ADMM_rho}
    \boldsymbol{\rho}^{(k+1)} = \phi\Big( \big(\boldsymbol{\rho}^{(i)}, \x^{(i+1)}, \z^{(i+1)}, \y^{(u+1)} \big)_{i=0}^k \Big), 
\end{equation} 
$\phi: \mathbb{R}^J \times \mathbb{R}^M \times \mathbb{R}^N \times \mathbb{R}^P \times \dots \to \mathbb{R}^J$
that selects new penalty parameters based on all current and past penalty parameters,
all current and past variables,
and, implicitly, the problem definition $f$, $g$, $\{\A_j\}_{j=1}^J$, $\{\B_j\}_{j=1}^J$, and $\{\c_j\}_{j=1}^J$.

The remainder of this paper is structured as follows. 
In Section \ref{sec:framework}, a framework for analyzing multiparameter ADMM for problems with multiple constraints as an affine fixed-point iteration on the dual variable is formulated. In Section \ref{sec:proposed_method}, this framework is used to derive the proposed multiparameter selection method by minimizing the spectral radius of the affine iteration matrix. In Section \ref{sec:existing_methods}, a brief survey of existing adaptive single-parameter methods and an additional multiparameter method is conducted, with these methods serving as comparison methods.  Section \ref{sec:problem_transformations} identifies an important class of problem transformations and characterizes the behavior of the proposed method and the comparison methods with respect to these problem transformations.  
In Section \ref{sec:expirements}, 
three numerical studies are conducted to demonstrate the benefits of the proposed method and compare it to the reference methods. Section \ref{sec:conclusion} presents the conclusions drawn from these results and possibilities for future extensions.

\section{Penalty Parameter Selection Framework}\label{sec:framework}

This section presents a framework for selection of multiple \ac{ADMM} penalty parameters. The presented framework is a multiparameter generalization of the single parameter spectral radius approximation (SRA) method \cite{mccann_robust_2024}.
The fundamental idea of this framework is to formulate \ac{ADMM} iterations locally as an affine fixed-point iteration, $\y^{(k+1)} = \afH \y^{(k)} + \afh $, where $\afH \in \mathbb{R}^{P\times P}$ and $\afh \in \mathbb{R}^{P}$ are dependent on penalty parameter selection. The theory of affine fixed-point iteration dictates that the fastest convergence is achieved when the spectral radius  of $\afH$ is minimized. Based on this analysis, the proposed method attempts to minimize the spectral radius of $\afH$. (While similar concepts underpin other approaches for ADMM parameter selection \cite{xu_adaptive_2017, lorenz_non_2019}, the resulting algorithms are different.)

In addition to extending the derivation of \cite{mccann_robust_2024} to the multiparameter case,
this work seeks to address one of its weakness. Specifically,  \cite{mccann_robust_2024} assumes that the iteration matrix $\afH$ has real eigenvalues for every $\boldsymbol{\rho}$ and that $\y^{(k+1)} - \y^{(k)}$ will converge to a single dominant eigenvector. This assumption is explicitly shown to be false via counter example in Section \ref{sec:expirements}. This work corrects the derivation resulting from this incorrect assumption and demonstrates that the original single parameter SRA method and proposed multiparameter method roughly achieve optimal performance.

\subsection{Iteration on $\y$}
\label{sec:itery}

This section provides a brief formulation of the \ac{ADMM} iterations in Eq. \eqref{eq:Multiconstraint_ADMM_x}-\eqref{eq:Multiconstraint_ADMM_yj} as an affine fixed-point iteration solely in terms of the dual variable or multiplier $\y$. This adapts the in-depth derivation in  \cite{mccann_robust_2024} for the multiparameter case, and is often expressed in literature  as Douglas Rachford splitting (DRS)~\cite{eckstein_douglasrachford_1992} on the dual problem  \cite{yan_self_2016, xu_adaptive_2017, lorenz_non_2019}. 

Applying first-order optimality conditions in terms of subgradients on the $\z$-update \cite{nesterov_introductory_2004, boyd_distributed_2011} 
$$\begin{array}{l}
\vec{0} \in \partial_{\z} g(\z^{(k+1)}) = \partial_{\z} g(\z^{(k+1)})  + \B^T \y^{(k+1)},
\end{array}$$ where $\partial_{\z} g(\z^{(k+1)})$ represents the subgradient of $g$ evaluated at $\z^{(k+1)}$. This then allows $\z^{(k)}$ to be expressed as a function of $\y^{(k)}$\begin{equation} \label{eq:G}
    \z^{(k)} = G \big(\y^{(k)} \big)  \quad 
    G(\w) = \argmin_\z g(\z) + \big(\B^T \w \big)^T \z ,
\end{equation} where $\w \in \mathbb{R}^P$.

Similarly, first-order optimality conditions in terms of subgradients can be applied for the $\x$-updates $$
\vec{0} \in \partial_\x f(\x^{(k+1)}) = \partial_\x f(\x^{(k+1)})  + \A^T \Tilde{\y}^{(k+1)}, $$ where $\partial_\x f(\x^{(k+1)}$ represents the subgradient of $f$ evaluated at $\x^{(k+1)}$ and  
 \begin{equation}\label{eq:synthetic_multiplier}
     \Tilde{\y}^{(k+1)} = \y^{(k)} + \D_{\boldsymbol{\rho}}( \A \x^{(k+1)} + \B \z^{(k)} - \c),
 \end{equation} is an introduced \emph{synthetic multiplier} or \emph{intermediate multiplier}. Then $\x^{(k)}$ can be expressed as a function of $\Tilde{\y}^{(k)}$ \begin{equation} \label{eq:F}
    \x^{(k)} = F\big(\tilde{\y}^{(k)}\big), \quad
    F(\w) = \argmin_{\x} f(\x) + \big(\A^T \w \big)^T \! \x .
\end{equation}

Using these definitions of $F$ and $G$, the \ac{ADMM} updates can be expressed solely in terms of the implicit multiplier updates.\begin{equation}\label{eqn:implicit_ytil_update}
 (\vec{I} - \D_{\boldsymbol{\rho}} \A F)(\tilde{\y}^{(k+1)}) = (\vec{I} + \D_{\boldsymbol{\rho}} \B G)(\y^{(k)}) - \D_{\boldsymbol{\rho}} \c. 
\end{equation}
\begin{equation}\label{eqn:implicit_y_update}
    (\vec{I} - \D_{\boldsymbol{\rho}} \B G)(\y^{(k+1)})  = \tilde{\y}^{(k+1)}  - \D_{\boldsymbol{\rho}} \B G(\y^{(k)}).
\end{equation}
Assuming $(\vec{I} - \D_{\boldsymbol{\rho}} \A F)$ and $(\vec{I} - \D_{\boldsymbol{\rho}} \B G)$ are locally invertible, the implicit updates in Eq. \eqref{eqn:implicit_ytil_update} and Eq. \eqref{eqn:implicit_y_update} can be summarized as

$$\tilde{\y}^{(k\!+\!1)} = (\vec{I} - \D_{\boldsymbol{\rho}} \A F)^{\!-\!1} \left( (\vec{I} + \D_{\boldsymbol{\rho}} \B G)(\y^{(k)}) - \D_{\boldsymbol{\rho}} \c\right), $$ $$\y^{(k\!+\!1)} = (\vec{I} - \D_{\boldsymbol{\rho}} \B G)^{\!-\!1}\left( \tilde{\y}^{(k+1)}  - \D_{\boldsymbol{\rho}} \B G(\y^{(k)})\right).$$

Note that $\tilde{\y}^{(k+1)}$ can be written purely in terms of $\y^{(k)}$ and merely acts as a placeholder variable, to improve readability.

\subsection{Affine Fixed-Point Iteration}
\label{sec:affinefixpt}

This section considers the case when ADMM is applied to a quadratic optimization problem. Using the logic applied in Section \ref{sec:itery}, this then allows \ac{ADMM} to be expressed as an affine fixed-point iteration and is a critical component to derive the proposed penalty parameter selection rule. Furthermore, it is assumed that convex functions can be locally well-approximated with a series of quadratics \cite{boggs_sequential_1995, rosen_convex_2004, lee_modified_2016, azagra_global_2013}, such as the local approximations utilized in Newton methods, \cite{dennis_quasi_1977, lewis_nonsmooth_2013}, thereby allowing the proposed method to be generalized to the broader class of convex optimization problems.  

Consider the case when $f$ and $g$ are of the forms  
$$f(\x) = \frac{1}{2}\x^T \Q \x + \q^T\x \quad   g(\z) = \frac{1}{2}\z^T \R \z + \r^T\z,$$
where $\Q$ is a positive definite $M\times M$ matrix, $\q \in \mathbb{R}^M$, $\R$ is a positive definite $N\times N$ matrix, and $\r \in \mathbb{R}^N$.

These definitions lead to a fixed point iteration in terms of the multiplier variable
$$\y^{(k+1)} = $$
$$  \underbrace{((\vec{I} + \D_{\boldsymbol{\rho}} \GG)^{\!-\!1} (\vec{I} + \D_{\boldsymbol{\rho}} \FF)^{\!-\!1} (\vec{I} + \D_{\boldsymbol{\rho}} \FF  \D_{\boldsymbol{\rho}} \GG)}_{\afH}\y^{(k)}  + 
\afh$$ where $\FF =  \A\Q^{-\!1}\A^T$, $\GG =  \B\R^{-\!1}\B^T$,  $\afH \in \mathbb{R}^{P\times P}$ is the update matrix that is dependent on $\boldsymbol{\rho}$, and $\afh \in \mathbb{R}^P$ is  an affine component. This fixed point iteration will be convergent when the spectral radius $\operatorname{rad}(\afH) <1. $

Supposing that these iterations converge to a unique fixed point $\y^*$, denote the error term $\eps^{(k)} = \y^{(k)} - \y^*$.  In the cases when $\afH$ only has real eigenvalues, $\eps^{(k)}$ will converge to a dominant eigenvector of $\afH$, as the components corresponding to smaller eigenvalue will rapidly vanish. Therefore, for sufficiently large $k$ and $\Delta k > 0 $ \begin{align}
\y^{(k + \Delta k)} - \y^{(k)} 
&\approx \big(r(\afH)^{\Delta k} - 1 \big) \eps^{(k)} ,
\end{align} that is $\y^{(k + \Delta k)} - \y^{(k)}$ will also be a maximal eigenvector.

However, in the cases when $\afH$ has  dominant eigenvalues that are complex, $\eps^{(k)}$ will instead converge to a real combination of the dominant eigenvectors $\v_{\boldsymbol{\rho}}$ and  $\overline{\v_{\boldsymbol{\rho}}}$, since the other eigenvector components will vanish rapidly. Let $\lambda_{\boldsymbol{\rho}}$ be one of the dominant eigenvalues, $\v_{\boldsymbol{\rho}}$ its corresponding eigenvector, and $(\overline{\lambda_{\boldsymbol{\rho}}}, \overline{\v_{\boldsymbol{\rho}}})$ their corresponding conjugate pair. Then for some complex coefficient $\zeta$ $$\eps^{(k)} \approx \frac{\zeta}{2} \v_{\boldsymbol{\rho}} + \overline{\frac{\zeta}{2} \v_{\boldsymbol{\rho}}} .$$

Therefore, for sufficiently large $k$ and $\Delta k > 0$ $$\y^{(k + \Delta k)} - \y^{(k)} \approx   \eps^{(k + \Delta k)} - \eps^{(k)} $$ $$ =  \frac{\zeta}{2} (\lambda_{\boldsymbol{\rho}}^{\Delta k } -1)   \v_{\boldsymbol{\rho}} + \overline{\frac{\zeta}{2}(\lambda_{\boldsymbol{\rho}}^{\Delta k } -1)\v_{\boldsymbol{\rho}}} .$$ This means that $\y^{(k + \Delta k)} - \y^{(k)}$ may not be a dominant eigenvalue, but is  contained within plane spanned by the real and imaginary components of the dominant eigenvector.

\section{Proposed Penalty Parameter Selection Method}
\label{sec:proposed_method}

This section presents the proposed selection method for multiple adaptive penalty parameter methods. The proposed selection rule is derived based on the affine fixed-point iteration introduced in the previous section,
and attempts to minimize the spectral radius of the iteration matrix by avoiding two limiting cases that are shown not to include the optimal penalty parameters.

Let $(\lambda_{\boldsymbol{\rho}}, \v_{\boldsymbol{\rho}})$ be a dominant eigenpair of $\afH$. Then $$(\vec{I} + \D_{\boldsymbol{\rho}} \GG)^{\!-\!1}(\vec{I} + \D_{\boldsymbol{\rho}} \FF)^{\!-\!1}(\vec{I} + \D_{\boldsymbol{\rho}} \FF \D_{\boldsymbol{\rho}} \GG)\v_{\boldsymbol{\rho}} = \lambda_{\boldsymbol{\rho}} \v_{\boldsymbol{\rho}}$$
$$\implies (\vec{I} + \D_{\boldsymbol{\rho}} \FF \D_{\boldsymbol{\rho}} \GG)\v_{\boldsymbol{\rho}} = \lambda_{\boldsymbol{\rho}} (\vec{I} + \D_{\boldsymbol{\rho}} \FF) (\vec{I} + \D_{\boldsymbol{\rho}} \GG)\v_{\boldsymbol{\rho}}.$$

Then $$|\lambda_{\boldsymbol{\rho}}|^2 = \frac{\|(\vec{I} + \D_{\boldsymbol{\rho}} \FF \D_{\boldsymbol{\rho}} \GG)\v_{\boldsymbol{\rho}} \|^2 }{\|(\vec{I} + \D_{\boldsymbol{\rho}} \FF) (\vec{I} + \D_{\boldsymbol{\rho}} \GG)\v_{\boldsymbol{\rho}}\|^2}.$$

\subsection{Dominating Cases} 

Consider the two cases when either $\v_{\boldsymbol{\rho}}$ dominates $\D_{\boldsymbol{\rho}}\GG\v_{\boldsymbol{\rho}}$ or vice versa. This section demonstrates that the optimal $\boldsymbol{\rho}$ does not satisfy either of these cases, and then proposes a simple and robust adaptive selection method for multiple penalty parameters by avoiding these two worst cases. This is demonstrated via a bounding argument that removes the need to linearize about the maximal eigenvector, as required in \cite{mccann_robust_2024}.

 \label{sec:proposed_method_ratio}

  \underline{\textbf{Case 1: $ \|\v_{\boldsymbol{\rho}} \| \gg \|\D_{\boldsymbol{\rho}}\GG\v_{\boldsymbol{\rho}} \| $.}}
Then \[|\lambda_{\boldsymbol{\rho}}|^2 \approx \frac{\|\v_{\boldsymbol{\rho}} \|^2 }{\|(\vec{I} + \D_{\boldsymbol{\rho}} \FF)\v_{\boldsymbol{\rho}}\|^2}.\] The eigenvalue norm can then be bounded above and below as $$\frac{1}{(1+\sigma_{\max}(\D_{\boldsymbol{\rho}} \FF))^2} \leq |\lambda_{\boldsymbol{\rho}}|^2 \leq \frac{1}{(1+\sigma_{\min}(\D_{\boldsymbol{\rho}} \FF))^2},$$ where $\sigma_{\min}(\D_{\boldsymbol{\rho}} \FF)$  and $\sigma_{\max}(\D_{\boldsymbol{\rho}} \FF)$ are the minimum and maximum singular values  of $\D_{\boldsymbol{\rho}} \FF.$ Both of these functions are strictly elementwise decreasing with respect to $\boldsymbol{\rho}.$ This implies that the global min for the eigenvalue norm is not achieved in \textbf{Case 1}.

Note that in \textbf{Case 1},$$(\vec{I} + \D_{\boldsymbol{\rho}} \FF \D_{\boldsymbol{\rho}} \GG)\v_{\boldsymbol{\rho}} = \lambda_{\boldsymbol{\rho}} (\vec{I} + \D_{\boldsymbol{\rho}} \FF) (\vec{I} + \D_{\boldsymbol{\rho}} \GG)\v_{\boldsymbol{\rho}} $$ $$\implies \v_{\boldsymbol{\rho}} \approx  (\vec{I} + \D_{\boldsymbol{\rho}} \FF) \lambda_{\boldsymbol{\rho}}\v_{\boldsymbol{\rho}} $$ which implies that $( \lambda_{\boldsymbol{\rho}}, \v_{\boldsymbol{\rho}})$ approaches an eigenpair of $(\vec{I} + \D_{\boldsymbol{\rho}} \FF)^{\!-\!1}$ and the eigenvalue $\lambda_{\boldsymbol{\rho}}$ and eigenvector $\v_{\boldsymbol{\rho}}$ are real.  

  \underline{\textbf{Case 2: $ \|\v_{\boldsymbol{\rho}} \| \ll \|\D_{\boldsymbol{\rho}}\GG\v_{\boldsymbol{\rho}} \| $.}}
  Then  $$|\lambda_{\boldsymbol{\rho}}|^2 \approx  \frac{\| \D_{\boldsymbol{\rho}} \FF \D_{\boldsymbol{\rho}} \GG\v_{\boldsymbol{\rho}} \|^2 }{\|(\vec{I} + \D_{\boldsymbol{\rho}} \FF)  \D_{\boldsymbol{\rho}} \GG\v_{\boldsymbol{\rho}}\|^2}.$$ The eigenvalue norm can then be bounded above and below by  $$ \left(\frac{\sigma_{\min}(\FF \D_{\boldsymbol{\rho}} \GG) }{(\sigma_{\max}(\GG) +  \sigma_{\min}(\FF \D_{\boldsymbol{\rho}} \GG) }\right)^2 \leq |\lambda_{\boldsymbol{\rho}}|^2 $$ $$ \leq \left( \frac{\sigma_{\max}(\FF \D_{\boldsymbol{\rho}} \GG) }{\sigma_{\min}(\GG) +  \sigma_{\max}(\FF \D_{\boldsymbol{\rho}} \GG) }\right)^2.$$ Both of these functions are strictly elementwise increasing with respect to $\boldsymbol{\rho}.$ This implies that the global minimum for the eigenvalue norm is not achieved in \textbf{Case 2}.

  Note that in \textbf{Case 2}, $$(\vec{I} + \D_{\boldsymbol{\rho}} \FF \D_{\boldsymbol{\rho}} \GG)\v_{\boldsymbol{\rho}} = \lambda_{\boldsymbol{\rho}} (\vec{I} + \D_{\boldsymbol{\rho}} \FF) (\vec{I} + \D_{\boldsymbol{\rho}} \GG)\v_{\boldsymbol{\rho}} $$ $$ \implies  \D_{\boldsymbol{\rho}} \FF \D_{\boldsymbol{\rho}} \GG\v_{\boldsymbol{\rho}} \approx (\vec{I} + \D_{\boldsymbol{\rho}} \FF)  \lambda_{\boldsymbol{\rho}}  \D_{\boldsymbol{\rho}} \GG\v_{\boldsymbol{\rho}}$$ which implies that $( \lambda_{\boldsymbol{\rho}},  \D_{\boldsymbol{\rho}} \GG\v_{\boldsymbol{\rho}})$ approaches an eigenpair of $(\vec{I} + \D_{\boldsymbol{\rho}} \FF)^{\!-\!1}\D_{\boldsymbol{\rho}} \FF$ and the eigenvalue $\lambda_{\boldsymbol{\rho}}$ and  $\D_{\boldsymbol{\rho}} \GG\v_{\boldsymbol{\rho}}$ are real.  $\D_{\boldsymbol{\rho}} \GG\v_{\boldsymbol{\rho}}$  being real similarly implies $\v_{\boldsymbol{\rho}}$ is because $\D_{\boldsymbol{\rho}}$ and  $\GG\v_{\boldsymbol{\rho}}$ are real. 

  \subsection{Proposed Penalty Parameter Selection Method} 
The proposed method is based on avoiding either of the previously mentioned \textbf{Case 1} and \textbf{Case 2}, that is $$ \|\v_{\boldsymbol{\rho}} \| \gg \|\D_{\boldsymbol{\rho}}\GG\v_{\boldsymbol{\rho}} \|     \quad \text{or} \quad\|\v_{\boldsymbol{\rho}} \| \ll \|\D_{\boldsymbol{\rho}}\GG\v_{\boldsymbol{\rho}} \|.$$ As noted, \textbf{Case 1} and \textbf{Case 2} both imply that both the eigenvalues and eigenvectors are real, or dominated by their real component. This means both cases can be described in terms of inequalities that only consider real components of the eigenvector $$\begin{array}{c}\|\operatorname{Real}(\zeta \v_{\boldsymbol{\rho}})\| \gg \|\D_{\boldsymbol{\rho}}\GG\operatorname{Real}(\zeta \v_{\boldsymbol{\rho}})\|     \quad \text{or} \\ \| \operatorname{Real}(\zeta \v_{\boldsymbol{\rho}} )\| \ll \|\D_{\boldsymbol{\rho}}\GG\operatorname{Real}(\zeta \v_{\boldsymbol{\rho}}) \|
\end{array},$$ for any chosen complex coefficient $\zeta$.

This can easily be avoided when the left and right sides are set to be equal $$\begin{array}{c} \|\operatorname{Real}(\zeta \v_{\boldsymbol{\rho}})\|^2 = \sum_{j=1}^J  \|\operatorname{Real}(\zeta \v_{\boldsymbol{\rho}})_j\|^2  \\ = \|\D_{\boldsymbol{\rho}}\GG\operatorname{Real}(\zeta \v_{\boldsymbol{\rho}}) \|^2 = \sum_{j=1}^J\rho_j^2  \| (\GG\operatorname{Real}(\zeta \v_{\boldsymbol{\rho}}))_j \|^2\end{array},$$ where $(\cdot)_j$ selects the components of $\v_{\boldsymbol{\rho}}$ and $\GG\v_{\boldsymbol{\rho}}$ corresponding to $\y_j$. This above equality is achieved when
$$ \|\operatorname{Real}(\zeta \v_{\boldsymbol{\rho}})_j\| = \rho_j   \| (\GG\operatorname{Real}(\zeta \v_{\boldsymbol{\rho}}))_j \|$$ holds for a chosen $\zeta$.

Choosing $\zeta $ such that $\y^{(k+1)}-\y^{(k)}\approx \frac{\zeta}{2} \v_{\boldsymbol{\rho}} + \overline{\frac{\zeta}{2} \v_{\boldsymbol{\rho}} } = \operatorname{Real}(\zeta \v_{\boldsymbol{\rho}})$, gives rise to the proposed multiparameter spectral radius approximation (MpSRA) rule \begin{equation} \label{eq:proposed_rho}
   \left(\rho_j^{(k+1)}\right)_\text{MpSRA} = \frac{ \big\| \y^{(k+1)}_j - \y^{(k)}_j \big\|}{\big\| \B_j (\z^{(k + 1)} - \z^{(k)}) \big\|}.
\end{equation}

Note that the resulting adaptive rule for $\boldsymbol{\rho}$ does not explicitly require that the problem be quadratic, and only requires values for $\y, \B, $ and $\z$. In practice, it is then possible to apply the proposed method to any ADMM algorithm for solving arbitrary convex problems.

Furthermore, the adaptive penalty parameters proposed in \eqref{eq:proposed_rho} can be formulated for the multiblock case as  \begin{equation} \label{eq:proposed_rho_multiblock}
   \left(\rho_j^{(k+1)}\right)_\text{SRA} = \frac{ \big\| \y^{(k+1)}_j - \y^{(k)}_j \big\|}{\big\| \tilde{\B_j} (\z_j^{(k + 1)} - \z_j^{(k)}) \big\|}.
\end{equation} In this form the proposed penalty parameters are defined independently for each multiblock subproblem. For applications of consensus ADMM, the proposed multiparameter method is easily adapted and creates separate penalty parameters for each subproblem and is fully compatible with with distributed and asynchronous approaches.

\subsection{Implementation of Proposed Method}

Two key points must be addressed for practical implementation of the proposed  MpSRA rule in Eq. \eqref{eq:proposed_rho}. 

First, the proposed rule requires that $\y^{(k+1)} -\y^{(k)}$ approximates the largest eigenvector of $\afH$, or is approximately in the plane determined by the dominant eigenvector pair in the complex eigenvalue case. This approximation requires that $k$ be sufficiently large. In practice, this can be achieved by only applying the proposed rule every $T$ iterations. Based on trial and error, $T = 5$ demonstrates sufficient performance.

Second, the rule presented in Eq. \eqref{eq:proposed_rho} does not directly account for the cases when $\y^{(k+1)}_j = \y^{(k)}_j$ or $\B_j \z^{(k+1)} = \B_j\z^{(k)}$. Standard ADMM requires a finite and positive penalty parameter and, via the equivalence of standard ADMM and multiparameter ADMM as demonstrated in Appendix \ref{app:multi_equivalence}, each element of the multiparameter $\rho_j$ must also be finite positive.
In these cases, multiplicative scaling is applied in a manner similar to residual balancing method \cite{wohlberg_admm_2017}. That is $\|\y^{(k+1)}_j - \y^{(k)}_j\| = 0 $ and $\|\B_j( \z^{(k+1)} - \z^{(k)})\| > 0 $ indicates that the ADMM method is weighted too much towards the constraint and a greater emphasis can be put on the objective by decreasing $\rho_j$ by a chosen factor $\tau^{\text{decr}}$. Similarly, $\|\y^{(k+1)}_j - \y^{(k)}_j\| > 0 $ and $\|\B_j( \z^{(k+1)} - \z^{(k)})\| = 0 $ indicates that the ADMM method is weighted too much towards the objective and a greater emphasis can be put on the constraint by increasing $\rho_j$ by a chosen factor  $\tau^{\text{incr}}$. If $\|\y^{(k+1)}_j - \y^{(k)}_j\| = 0 $ and $\|\B_j( \z^{(k+1)} - \z^{(k)})\| = 0 $, both the constraint and objective are equally weighted and no adjustment to $\rho_j$ is needed. 
In practice, $\tau^{\text{decr}} = \tau^{\text{incr}}=10$ demonstrates sufficient performance. 

This practical implementation of the proposed rule in Eq. \eqref{eq:proposed_rho} is described in Algorithm \ref{algo:proposed}.

\begin{algorithm}
\linespread{1.2}\selectfont
\SetKwInput{KwParameters}{Parameters}
\SetKwIF{If}{ElseIf}{Else}{if}{:}{else if}{else}{endif}
\SetKw{And}{and}
\caption{Proposed multiparameter spectral radius approximation (MpSRA) $\boldsymbol{\rho}$ selection method}
\label{algo:proposed}
\KwIn{$k, \boldsymbol{\rho}^{(k)},  \z^{(k)}, \z^{(k+1)}, \y^{(k)}, \y^{(k+1)}$}
\KwParameters{$T=5, \tau^{\text{incr}}=\tau^{\text{decr}}=10$}
\KwOut{$\boldsymbol{\rho}^{(k+1)}$}
\uIf{$k \mod T \ne 0$}
{
    \Return $\boldsymbol{\rho}^{(k)}$
}

\For{$j = 1,\hdots, J$}{
$p_j \gets \big\|\y_j^{(k+1)} - \y_j^{(k)} \big\|$ \\
$q_j \gets \big\| \B_j \big( \z^{(k+1)} - \z^{(k)} \big) \big\|$ \\
\uIf{$p_j = 0$ \And $q_j > 0$}
{
 $\rho^{(k+1)}_j \leftarrow \rho_j^{(k)} / \tau^{\text{decr}}$
}
\uElseIf{$p_j > 0 $ \And $q_j = 0$}
{
$\rho^{(k+1)}_j \leftarrow \tau^{\text{incr}} \rho^{(k)}$
}
\uElseIf{$p_j = 0 $ \And $q_j = 0$}
{
$\rho^{(k+1)}_j \leftarrow \rho_j^{(k)} $
}
\uElse{
$\rho^{(k+1)}_j \leftarrow p_j/q_j$
}}

$\boldsymbol{\rho}^{(k+1)} \leftarrow (\rho_1^{(k+1)},\hdots,\rho_J^{(k+1)})^T$

\Return $\boldsymbol{\rho}^{(k+1)}$

\end{algorithm}

\section{Existing Penalty Parameter Methods}\label{sec:existing_methods}

This section briefly outlines and describes other state-of-the-art adaptive penalty parameter methods proposed in other works.

\subsection{Single-Parameter Methods}

Several works have proposed adaptive single-parameter methods for the single-constraint version of ADMM. We focus here on the same four methods analyzed in \cite{mccann_robust_2024}, which represent the state-of the-art methods in practice. These single-parameter methods are the residual balancing  (RB) method, proposed in  \cite{he-2000-alternating} and applied in \cite{hansson-2012-subspace, liu-2013-nuclear, vu-2013-fantope, iordache-2014-collaborative, weller-2014-phase, wohlberg-2014-efficient},  the spectral radius bound (SRB) method \cite{lorenz_non_2019}, the single-parameter  Barzilai-Borwein spectral (BBS) method \cite{xu_adaptive_2017, xu_adaptive_relaxed_2017, barzilai1988two}, and the single-parameter spectral radius approximation (SRA) method \cite{mccann_robust_2024}.

\subsection{Multiparameter Barzilai-Borwein Spectral Method}

Although many single-parameter methods can sometimes be adapted to multiparameter analogs for multiconstraint problems, few works have implemented multiparameter versions of ADMM. One exception is the multiparameter version of the Barzilai-Borwein spectral (MpBBS) method \cite{xu_alternating_2019}. The multiparameter BBS method serves as the main point of comparison for the proposed multiparameter method. 

The multiparameter BBS method \cite[Section 5.3.3]{xu_alternating_2019} is the natural extension of the single-parameter BBS method and is derived in a similar manner. The multiparameter BBS method is expressed. \begin{equation}\label{eq:multi_BBS}
\begin{array}{l}
     \left(\rho_j^{(k+1)}\right)_\text{MpBBS} =   \\ \\ 
          \sqrt{\frac{ \big\| \Delta \tilde{\y}^{(k)}_j \big\|^2 
            \big\| \Delta \y^{(k)}_j \big\|^2
        }{\ip{\A_j(\Delta \x^{(k)})
            }{ \Delta \tilde{\y}^{(k)}_j } 
       \ip{ \B_j(\Delta \z^{(k)}) }{\Delta \y^{(k)}_j}}}.\end{array}\end{equation}
       where $\Delta \x^{(k)} = \x^{(k)} - \x^{(k - \Delta k)}$, $\Delta \z^{(k)} = \z^{(k)} - \z^{(k - \Delta k)}, \Delta \y^{(k)}_j = \y^{(k)}_j - \y^{(k - \Delta k)}_j, $  $\tilde{\y}^{(k)}_j$ is defined the same as in \eqref{eq:synthetic_multiplier}, $\Delta \tilde{\y}^{(k)}_j = \tilde{\y}^{(k)}_j - \tilde{\y}^{(\Delta k)}_j$ and $\Delta k$ is some positive integer for delay in iterations. In practice, utilizing the BBS method requires a variety of additional safeguards based on a complicated heuristic assessment of curvature values derived in the formulation of the BBS method.

\section{Problem Multiscaling}\label{sec:problem_transformations}

This section identifies an important type of problem transformation for optimization problems with multiple constraints. This problem transformation, multiscaling, introduces a family of optimization problems whose solutions are related via a corresponding transformation in  the solution space. This section addresses how multiscaling alters the behaviors of each adaptive penalty parameter method and introduces the concept of multiscaling covariance, which is  a generalization of scaling covariance \cite{wohlberg_admm_2017, mccann_robust_2024}, intended to classify a family of optimization problems that share constraints at different scales. 

Consider using \ac{ADMM} to solve a member of a family of optimization problems \begin{equation} \label{eq:admm_problem_multi_scaled}
\begin{array}{ccc}
    \displaystyle{\min_{\x, \z}} & \alpha f(\gamma \x) + \alpha g(\delta \z) &  \\
   \text{s.t.}  &  \beta_j \A_j \gamma \x + \beta_j \B_j \delta \z = \beta_j \c_j & j = 1,\hdots,J
\end{array}
\end{equation} where the family is parameterized by the scalars $\alpha, \{\beta_j\}_{j=1}^J, \gamma,$ and $\delta$. We refer to the problem with $\alpha = \beta_1 = \hdots = \beta_J = \gamma = \delta = 1$ as the \emph{unscaled} problem. It is important to note that there is no definitive choice of the unscaled problem and it merely serves as a reference member of the family of problems.

Denote $\x^*, \z^*$ as the solution to the unscaled problem. Then the solution to the problem with scaling $\alpha,\{\beta_j\}_{j=1}^J,\gamma,\delta$ is \begin{equation}
\bar{\x}^*=\frac{\x^*}{\gamma}, \qquad \bar{\z}^*=\frac{\z^*}{\delta} ,
\end{equation} and is independent of choice of $\alpha$ or $\{\beta_j\}_{j=1}^J$. 

However, note that when multiconstraint ADMM is implemented for the unscaled problem with a penalty parameter $\boldsymbol{\rho} = (\rho_1,\hdots,\rho_J)^T$, equivalent behavior in ADMM for the $(\alpha,\{\beta_j\}_{j=1}^J,\gamma,\delta)$-scaled problem is demonstrated with the scaled penalty parameter $\overline{\boldsymbol{\rho}} = (\frac{\alpha}{\beta_1^2}\rho_1,\hdots, \frac{\alpha}{\beta_J^2} \rho_J)^T.$ This means that although scaling the constraints does not affect the solution to the optimization problem, it affects penalty parameter selection and thereby convergence behavior of ADMM. 

Ideally, then a multiconstraint ADMM penalty parameter method should scale properly with scaling applied to the optimization problem, motivating the following definition. 

\begin{definition}[Multiscaling Covariant]\label{def:multi_scaling_covariant}
A multiconstraint \ac{ADMM} penalty parameter selection method, $\phi$,
is \emph{multiscaling covariant} if
\begin{multline}\label{eq:scaling_covariance}
    \phi'\left( \left(\alpha \D_{\boldsymbol{\beta}}^{-2} \boldsymbol{\rho}^{(\ell)}, \frac{\x^{(\ell+1)}}{\gamma}, \frac{\z^{(\ell+1)}}{\delta}, \alpha \D_{\boldsymbol{\beta}}^{\!-\!1} \y^{(\ell+1)} \right)_{\ell=0}^k \right) \\
    =
    \alpha  \D_{\boldsymbol{\beta}}^{-2}
    \phi\left( \left(\boldsymbol{\rho}^{(\ell)}, \x^{(\ell+1)}, \z^{(\ell+1)}, \y^{(\ell+1)} \right)_{\ell=0}^k \right) ,
\end{multline} where $\boldsymbol{\beta} = (\beta_1,\hdots,\beta_J)^T$ and $\D$ is the diagonal matrix operator defined in Eq. \eqref{eq:diagonal_matrix_operator} and $\phi'$ is selection criteria for the scaled problem, i.e.,
where $\phi$ depends on $f$, $g$, $\A$, $\B$, and $\c$,
 $\phi'$ depends on $\alpha f(\gamma \cdot )$, $\alpha g(\delta \cdot )$, $\gamma \D_{\boldsymbol{\beta}}\A$, $\delta \D_{\boldsymbol{\beta}}\B$, and $\D_{\boldsymbol{\beta}}\c$.
\end{definition} 

That is, if a method is multiscaling covariant it will select $\boldsymbol{\rho}^{(k)}$ at iteration $k$
of the unscaled problem,
it will select $\alpha  \D_{\boldsymbol{\beta}}^{-2} \boldsymbol{\rho}^{(k)}$
for each corresponding scaled problem. Parameter selection methods being multiscaling covariant is critical because problem scaling is unavoidable in practice. For example, in many engineering problems, problem structure is determined by the choice of units of measurement, or scale, within the objective and constraints and an effective optimization needs to be independent of this choice.

Note that the characterized family of problems in Eq. \eqref{eq:admm_problem_multi_scaled} and Definition \ref{def:multi_scaling_covariant} can be further generalized by replacing the scalars $\gamma$ and $\delta$ with invertible matrices $\boldsymbol{\Gamma} \in \mathbb{R}^{M\times M}$ and $\boldsymbol{\varDelta} \in \mathbb{R}^{N\times N}$. In practice, members of this generalized family of problems can exhibit different behaviors in numerical solutions, potentially in cases when $\boldsymbol{\Gamma} $ or $\boldsymbol{\varDelta}$ are ill-conditioned. Reference \cite{mccann_robust_2024} provides additional analysis on adaptive methods being translation invariant. Note that MpSRA inherits translation invariance from the single parameter SRA.

\subsection{Single Parameter Selection Methods}

Consider an ADMM single penalty parameter selection method \begin{equation} \label{eq:single_adaptive_ADMM_rho}
    {\rho}^{(k+1)} = \psi\Big( \big({\rho}^{(\ell)}, \x^{(\ell+1)}, \z^{(\ell+1)}, \y^{(\ell+1)} \big)_{\ell=0}^k \Big). 
\end{equation} 

This update is defined in terms of a function
$\psi: \mathbb{R} \times \mathbb{R}^M \times \mathbb{R}^N \times \mathbb{R}^P \times \dots \to \mathbb{R}$. This can be viewed as an equivalent adaptive multiparameter rule that only inputs and outputs $\boldsymbol{\rho}^{(k)}$ such that $\rho_1^{(k)} = \hdots = \rho_J^{(k)}.$

A single-parameter method can only scale according to Eq. \eqref{eq:scaling_covariance} in the $\beta_1= \hdots = \beta_J$ case. This limitation means that a single-parameter method may be scaling covariant with respect to scaling by a single scalar but cannot be multiscaling covariant.

\subsection{Multiparameter BBS}
 The multiparameter BBS method is multiscaling covariant, since  for each $j$

$$\phi^{\text{MpBBS}'}_j\left( \left(\alpha \D_{\boldsymbol{\beta}}^{-2} \boldsymbol{\rho}^{(\ell )}, \frac{\x^{(\ell+1)}}{\gamma}, \frac{\z^{(\ell+1)}}{\delta}, \alpha \D_{\boldsymbol{\beta}}^{\!-\!1} \y^{(\ell +1)} \right)_{\ell=0}^k \right)$$

$$ = \sqrt{\frac{
            \big\| \frac{\alpha}{\beta_j} \Delta \tilde{\y}^{(k)}_j \big\|^2 
            \big\| \frac{\alpha}{\beta_j} \Delta \y^{(k)}_j \big\|^2
        }{
        \ip{
                 \beta_j\A_j\gamma (\frac{\Delta \x^{(k)}}{\gamma})
            }{
               \frac{\alpha}{\beta_j} \Delta \tilde{\y}^{(k)}_j
            } 
       \ip{
                \beta_j\B_j\delta (\Delta \frac{\z^{(k)}}{\delta})
            }{
               \frac{\alpha}{\beta_j} \Delta \y^{(k)}_j
            }
    }
    } .
$$

$$ = \frac{\alpha}{\beta_j^2} 
          \sqrt{
    \frac{
            \big\| \Delta \tilde{\y}^{(k)}_j \big\|^2 
            \big\| \Delta \y^{(k)}_j \big\|^2
        }{
        \ip{
                \A_j(\Delta \x^{(k)})
            }{
                \Delta \tilde{\y}^{(k)}_j
            } 
       \ip{
                \B_j(\Delta \z^{(k)})
            }{
                \Delta \y^{(k)}_j
            }
    }
    } .$$

    $$= \frac{\alpha}{\beta_j^2} \phi^{\text{MpBBS}}_j \left( \left(\boldsymbol{\rho}^{(\ell )}, \x^{(\ell+1)}, \frac{\z^{(\ell+1)}}{\delta}, \y^{(\ell +1)} \right)_{\ell=0}^k \right).$$

\subsection{Multiparameter SRA}

The proposed MpSRA method is multiscaling covariant since for each $j$
$$\phi^{\text{MpSRA}'}_j\left( \left(\alpha \D_{\boldsymbol{\beta}}^{-2} \boldsymbol{\rho}^{(\ell )}, \frac{\x^{(\ell+1)}}{\gamma}, \frac{\z^{(\ell+1)}}{\delta}, \alpha \D_{\boldsymbol{\beta}}^{\!-\!1} \y^{(\ell +1)} \right)_{\ell=0}^k \right)$$ $$ = \frac{\frac{\alpha}{|\beta_j|} \| \y_j^{(k+1)} - \y_j^{(k)}\| }{|\beta_j| \|\B_j \delta (\frac{\z^{(k+1)}}{\delta } - \frac{\z^{(k)}}{\delta})\| }$$
$$ =\frac{\alpha}{\beta_j^2} \frac{ \| \y_j^{(k+1)} - \y_j^{(k)}\|}{\|\B (\z^{(k+1)} - \z^{(k)})\| }$$
    $$= \frac{\alpha}{\beta_j^2} \phi^{\text{MpSRA}}_j \left( \left(\boldsymbol{\rho}^{(\ell )}, \x^{(\ell+1)}, \frac{\z^{(\ell+1)}}{\delta}, \y^{(\ell +1)} \right)_{\ell=0}^k \right).$$ 

\section{Experiments and Results}\label{sec:expirements}

We performed 
three numerical experiments to demonstrate the benefits of the proposed multiparameter method and compare it to other adaptive penalty parameter approaches. In each of these experiments, the proposed and comparison methods were applied to a different optimization. The first experiment applied these methods for solving a constrained sum of quadratics optimization problem that resulted in an iteration matrix with complex eigenvalues for some penalty parameter selections and assessed our analysis of the iteration matrix and eigenvalue behavior. The second experiment applied these methods for solving a constrained sum of quadratics optimization problem with multiple scales between constraints and assessed the ability of the proposed method to adjust to scales between constraints. 
The third experiment applied these methods to solving a multiblock formulation of an image reconstruction problem with $\ell_1$ data fidelity and TV regularization and assessed each method for a computationally expensive, non-quadratic problem.

\subsection{Sum of Quadratics Resulting in an Iteration Matrix with Complex Eigenvalues}

Consider the constrained optimization problem 
\begin{equation}\label{eq:complex_quadratic}
\begin{array}{cc}
     \displaystyle{\argmin_{\x,\z}} &  \frac{1}{2} \x^T \Q \x + \q^T \x + \frac{1}{2}\z^T\R\z + \r^T\x \\ 
    \text{s.t. } & \x + \z = \c
\end{array}, 
\end{equation} where $\x,\z \in \mathbb{R}^2$, $\R = \operatorname{diag}(0.1, 10)$, $\Q = \boldsymbol{U}_\theta \R \boldsymbol{U}_\theta^T$
$$\boldsymbol{U}_\theta  = \begin{pmatrix}
    \cos(\theta) & - \sin(\theta) \\ \sin(\theta) & \cos(\theta)
\end{pmatrix},$$ $\theta = \frac{\pi}{4}$, $\q = (1,1)^T, \r = (1, -1)^T$, and $\c = (2,1)^T.$ In this setting the constraint can be split into two subconstraints,  $\x_1 + \z_1 = \c_1$ and $\x_2 + \z_2 = \c_2$, and a multiconstraint ADMM formulation can be applied. The corresponding ADMM iteration matrix is
$$\afH = (I  + \D_{\boldsymbol{\rho}} \R^{\!-\!1})^{\!-\!1} (I  + \D_{\boldsymbol{\rho}} \Q^{\!-\!1})^{\!-\!1}(\vec{I} + \D_{\boldsymbol{\rho}}\Q^{\!-\!1}\D_{\boldsymbol{\rho}} \R^{\!-\!1}).$$

\begin{figure*}
    \centering
    \includegraphics[width=0.9\columnwidth]{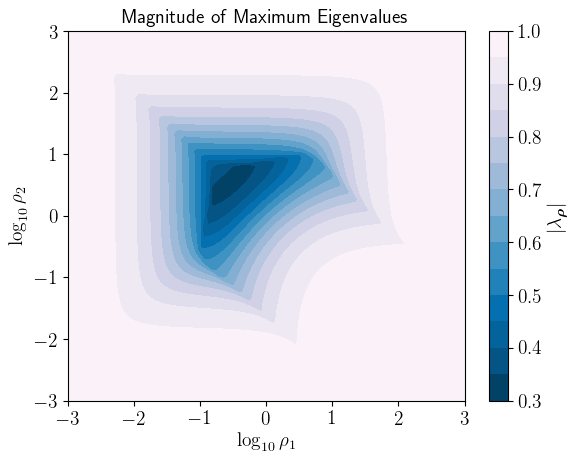} \includegraphics[width=0.9\columnwidth]{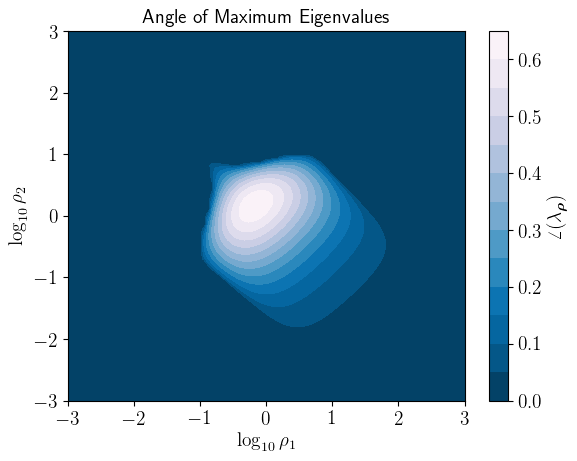} 
    \caption{Magnitude and angle (radians) of maximum eigenvalues of complex iteration matrix $\afH$ plotted as a function of $\rho_1$ and $\rho_2$. The maximum eigenvalues become real when either $\rho_1$ or $\rho_2$ are very large or very small. The the optimal fixed $\boldsymbol{\rho} = (\rho_1,\rho_2)^T$ corresponds to the location of the smallest magnitude, which occurs for a complex maximal eigenvalue. }
    \label{fig:eigen_example}
\end{figure*}

\begin{figure*}
    \centering
    \includegraphics[width=0.8\columnwidth]{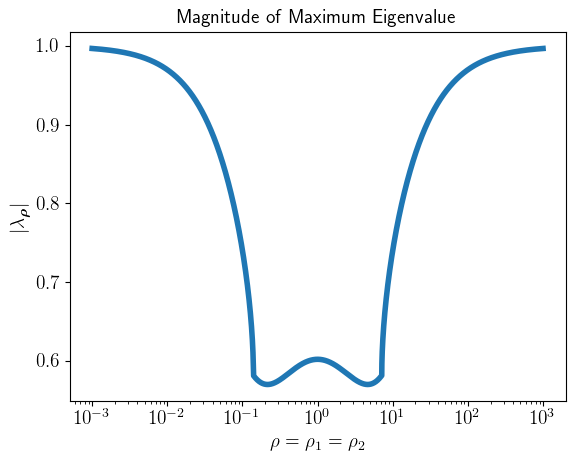} \ \ \ \ \ \ \ \ \ \includegraphics[width=0.8\columnwidth]{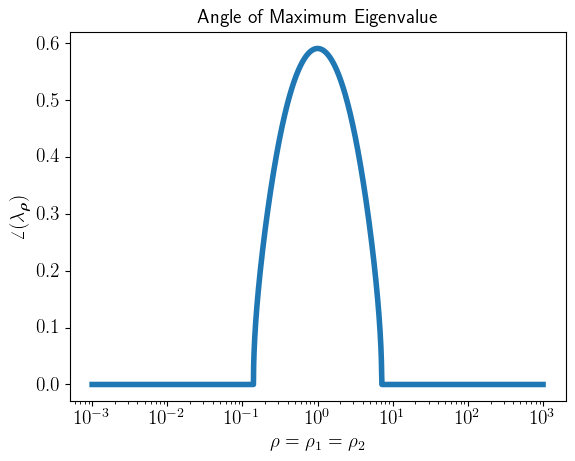} 
    \caption{Magnitude and angle (radians) of maximum eigenvalues of complex iteration matrix $\afH$ plotted as a function of $\rho = \rho_1 = \rho_2$. }
    \label{fig:diag_eigen_example}
    \vspace{-3mm}
\end{figure*}

Notably, $\afH$ will have complex eigenvalues for some values of $\boldsymbol{\rho} = (\rho_1,\rho_2)^T. $ Figure \ref{fig:eigen_example} displays surfaces for the magnitude and angle of the maximum eigenvalues of $\afH$. Similarly, the magnitude and angles of the maximum eigenvalues in the cases when $\rho = \rho_1 = \rho_2$, corresponding to the diagonals in Figure \ref{fig:eigen_example}, are plotted in Figure \ref{fig:diag_eigen_example}.

\begin{figure*}
    \centering
    \includegraphics[width=0.9\linewidth]{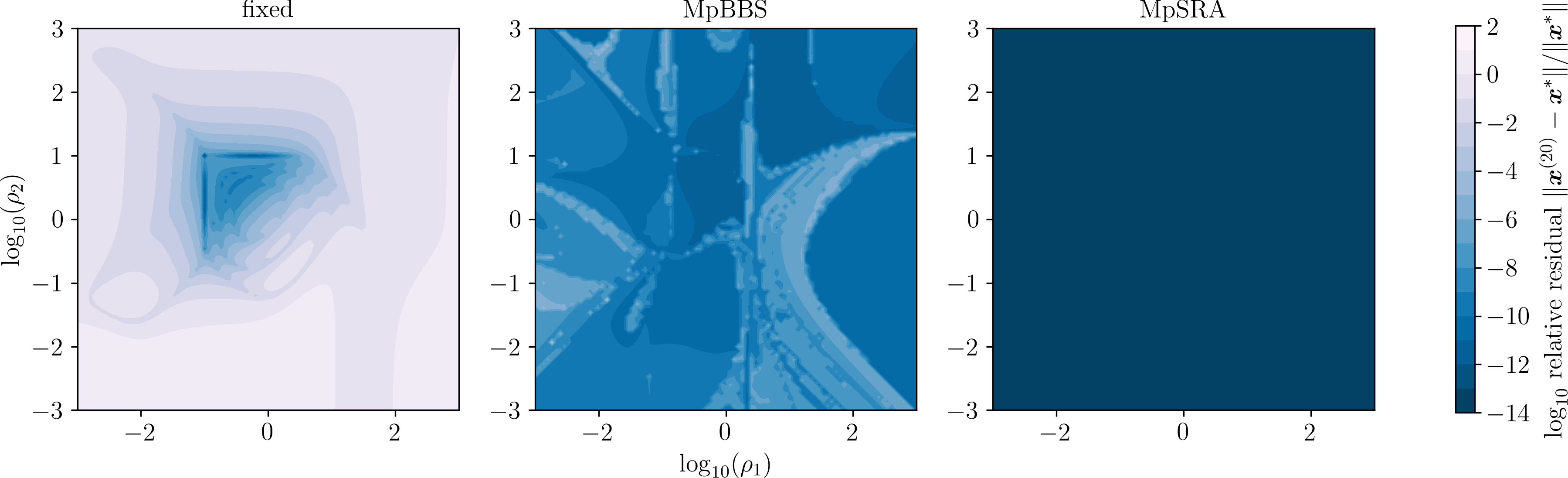}\\
    \vspace{3mm}
    \includegraphics[width=0.9\linewidth]{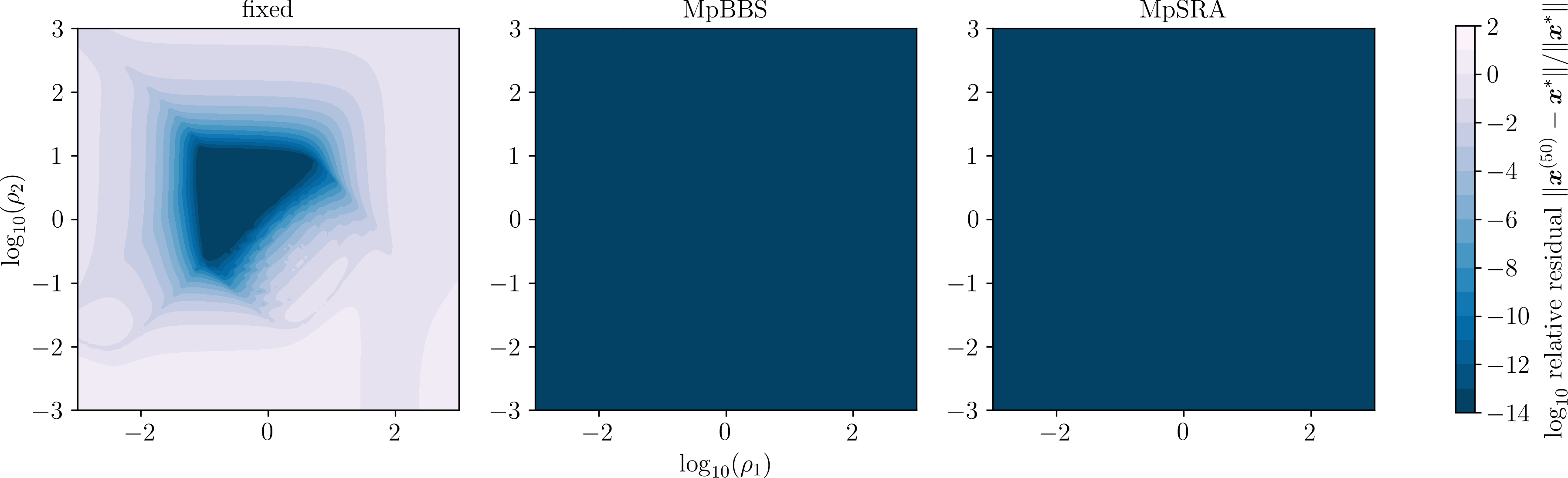}
    \caption{Relative residual of ADMM solutions corresponding to an iteration matrix with complex eigenvalues for fixed penalty parameter, multiparameter BBS, and multiparameter SRA methods after 20 and 50 iterations plotted on a surface as a function of initial $\rho_1$ and $\rho_2$. Note that the structure of the fixed method's residual plots mimics the eigenvalue structure in Fig. \ref{fig:eigen_example} and only converges by 50 iterations for a specific set of $\boldsymbol{\rho}$.  At 20 iterations the multiparameter SRA method has converged to a relative residual close to zero for the entire $\boldsymbol{\rho}$ search space, while the multiparameter BBS method requires 50 iterations. }
    \label{fig:worst_quad}
    \vspace{-8mm}
\end{figure*}

\begin{figure*}
\centering
\includegraphics[width=0.8\columnwidth]{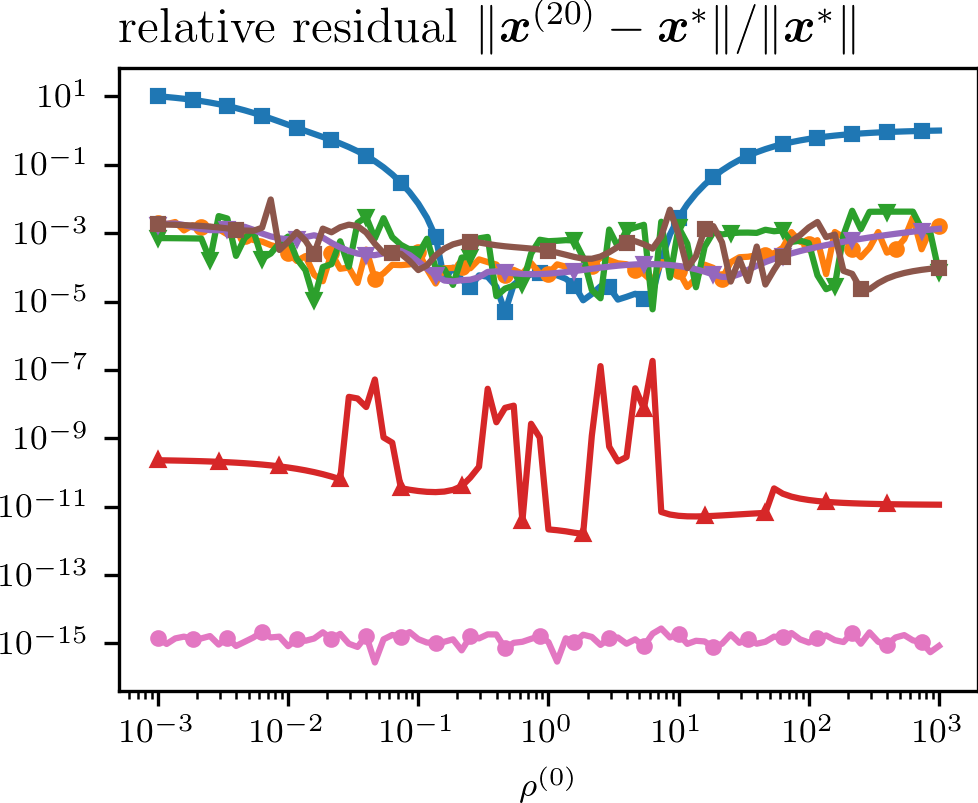} \ \ \includegraphics[width=0.8\columnwidth]{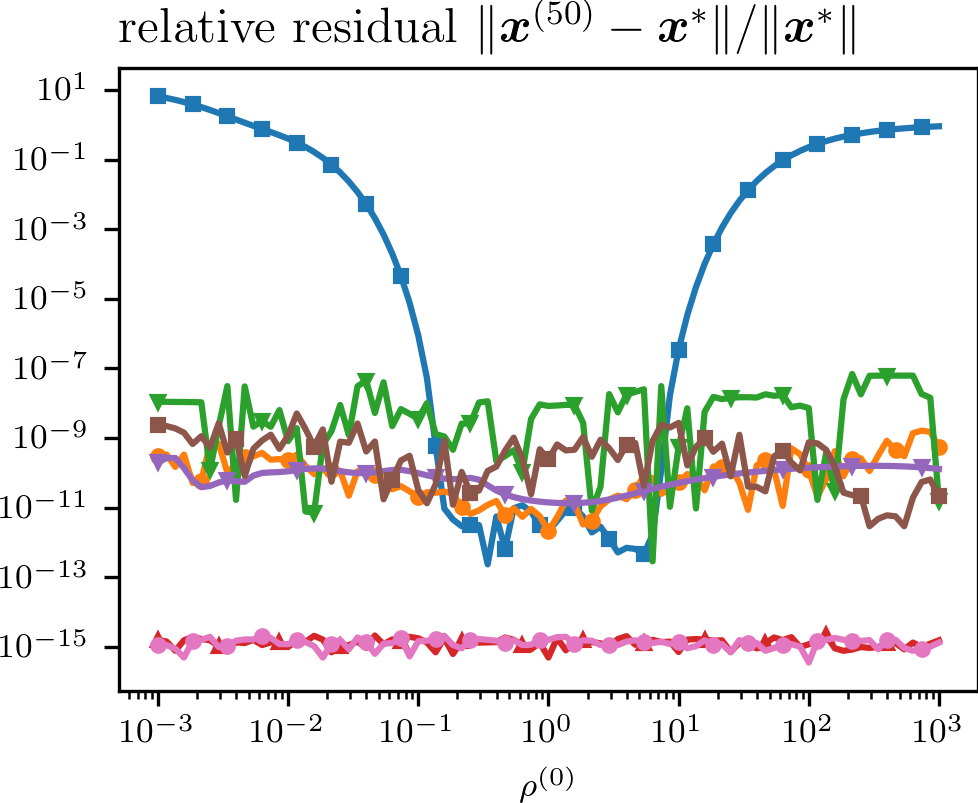}
\includegraphics[width=0.7\columnwidth, angle = 90]{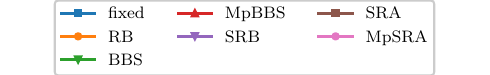}
\caption{Relative residual of ADMM solutions corresponding to an iteration matrix with complex eigenvalues for single-parameter and multiparameter adaptive penalty parameter rules after 20 and 50 iterations plotted as a function of initial $\rho = \rho_1 = \rho_2$.  Note that the structure of the fixed method's residual plots mimics the eigenvalue structure in Fig. \ref{fig:diag_eigen_example}. Both the multiparameter BBS and SRA method outperforms and converge quicker than the single-parameter methods, with the proposed multiparameter SRA method converging the quickest (20 iterations instead of 50).}
\label{fig:worst_quad1}
\end{figure*}

The relative residual of the multiconstraint algorithms for an array of starting $(\rho_1,\rho_2)$ after 20  and 50 iterations is displayed as a surface in Fig. \ref{fig:worst_quad}. The relative residuals of the single-constraint and multiconstraint algorithms for a range of starting $\rho = \rho_1 = \rho_2$ after 20 and 50 iterations are plotted in Fig. \ref{fig:worst_quad1}.


\subsection{Scaled Constraint Sum of Squares}

Consider the constrained optimization problem 
\begin{equation}\label{eq:scaled_sum_of_squares}
\begin{array}{cc}
     \displaystyle{\argmin_{\x,\z}} &  \frac{1}{2} \x^T \Q \x + \q^T \x + \frac{1}{2}\z^T\R\z + \r^T\x  \\ 
    \text{s.t. } & j^m(\vec{a}_j^T \x + \vec{b}_j^T\z - \c_j) = 0, \ j = 1,\hdots, J,
\end{array}
\end{equation} where each constraint is scaled by it index $j$ raised to a chosen power $m$, $\x \in \mathbb{R}^M, \z \in \mathbb{R}^N,$ $\vec{a}_j$ and $\q$ are randomly sampled from an $M$-dimensional standard normal distribution, $\vec{b}_j$ and $\r$ are randomly sampled from an $N$-dimensional standard normal distribution, $\Q = \Q_1^T\Q_1$, $\Q_1$ is randomly sampled from an $M\times M$ standard normal distribution, $\R = \R_1^T\R$, $\R_1$ is drawn from an $N\times N$ standard normal distribution, $\c_j$ are scalar variables drawn from a random normal distribution, and $m$ is a scaling parameter between the different constraints.

\begin{figure*}
    \centering
    \includegraphics[width=0.3 \textwidth]{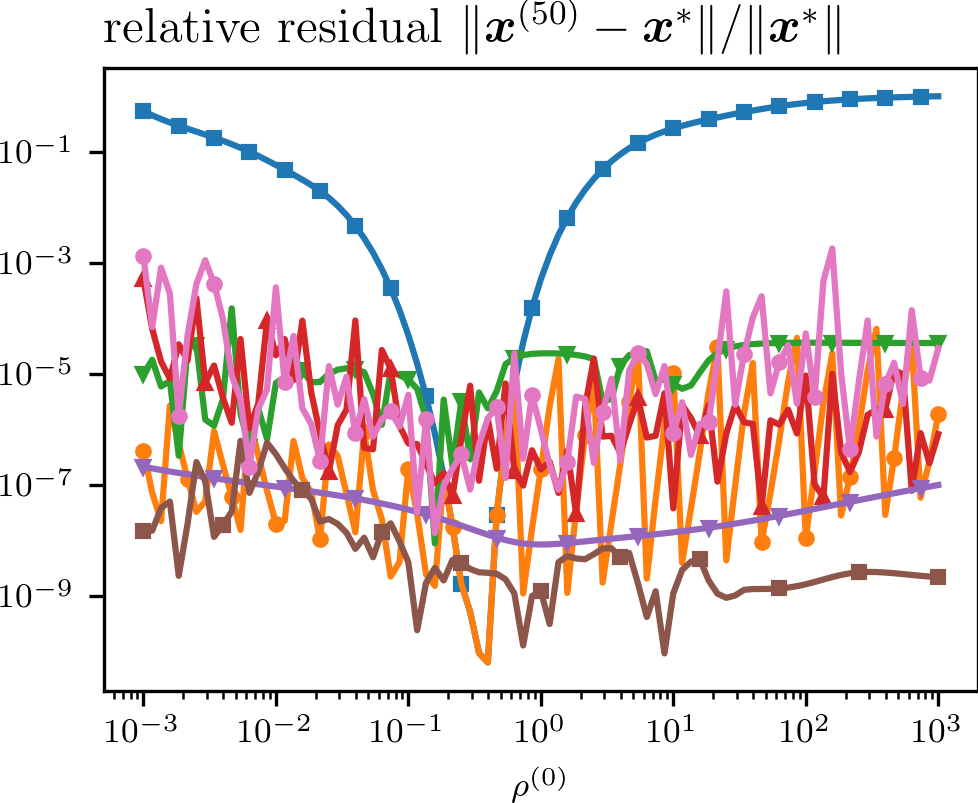}    \includegraphics[width=0.3 \textwidth]{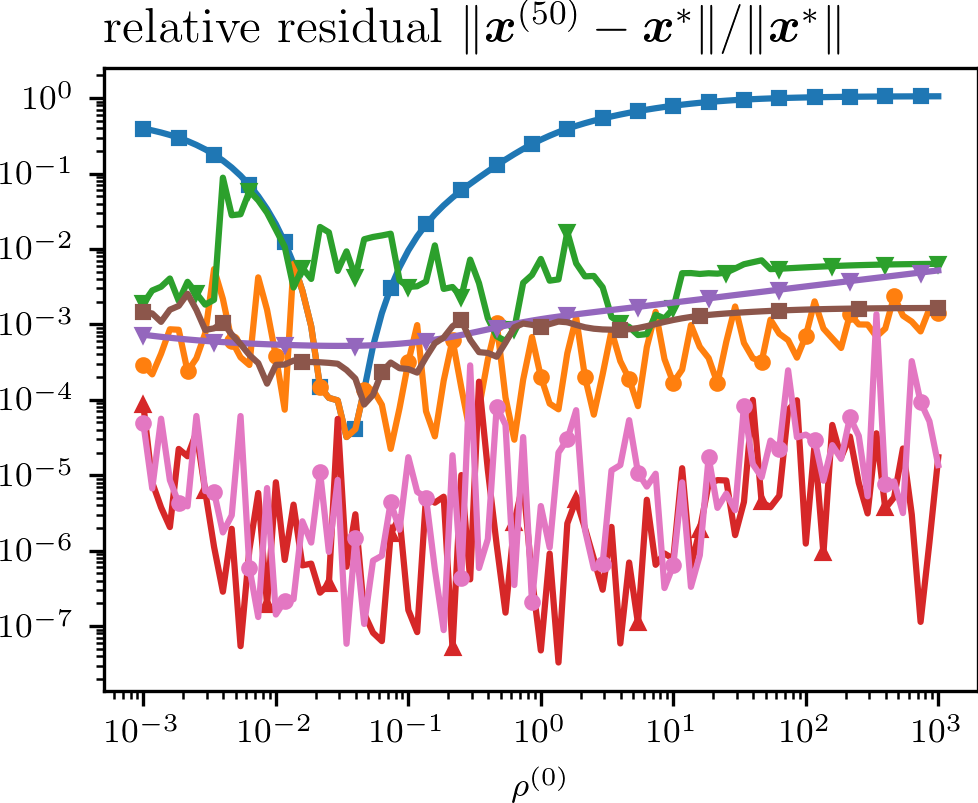}  \includegraphics[width=0.3 \textwidth]{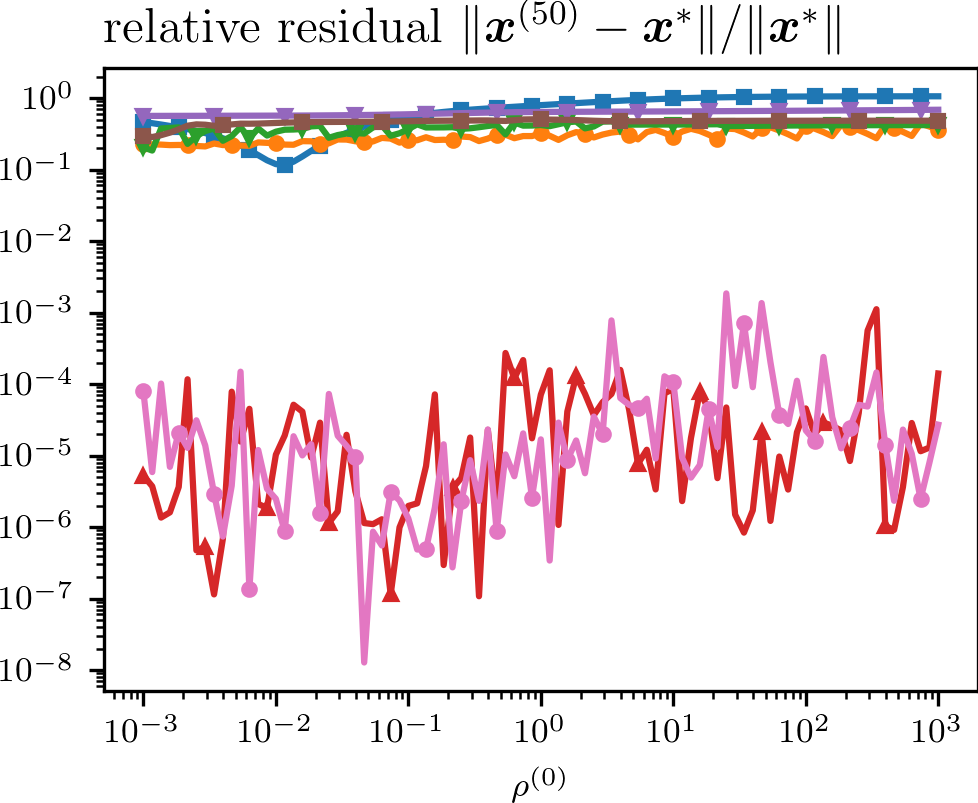}  \includegraphics[width=0.55\columnwidth, angle = 90]{Figures/multiblock_l1_df_total_variation_regularization_legend.pdf}\caption{Relative residual after 50 iterations for ADMM solutions of $m=0$ (left), $m=1$ (middle), and $m=2$ (right) sum of quadratics problem for each adaptive penalty parameter method plotted as a function of initial $\rho = \rho_1 = \rho_2$. The optimal $\rho$ shifts for each case of $m$ and the single $\rho$ methods perform worse as the scaling between constraints grows. The multiparameter methods do not perform the best at the $m=0$ case when the constraints are scaled evenly. However, the performance of the multiparameter methods demonstrate stable performance as the scaling between constraints grows.}
    \label{fig:sum_of_quadratics_multiconstraint}
    \vspace{-1mm}
\end{figure*}

The relative residuals of the single-constraint and multiconstraint algorithms for a range of starting $\rho = \rho_1 = \rho_2$ after 50 iterations for the $m=0,1,2$ cases are plotted in Fig. \ref{fig:sum_of_quadratics_multiconstraint}.

\subsection{Multiblock $\ell_1$ Data Fidelity Total Variation Regularization, Image Reconstruction}

Consider the unconstrained optimization problem for image reconstruction
\begin{equation}\label{eq:l1_tv_recon_unconstrained}
\begin{array}{cc}
     \displaystyle{\argmin_{\x}} &  \| \mathcal{F}(\x) - \d\|_1 + \delta \| \nabla \x \|_{2,1}
\end{array}, 
\end{equation}   where $\x \in \mathbb{R}^{M\times M}$ is the space of images to reconstruct over, $\mathcal{F}:\mathbb{R}^{M\times M} \rightarrow \mathbb{R}m$ represents a linear imaging operator that maps from an image to a set of $m$ measurements,  $\d\in\mathbb{R}^{m}$ represents a set of noisy measurements, $\|\cdot\|_1$ represents the $\ell_1$ norm over the $M\times M$ image, $\nabla:\mathbb{R}^{M\times M} \rightarrow \mathbb{R}^{2\times M\times M}$ is the finite difference gradient operator, $\|\cdot\|_{2,1}$ is the 2-dimensional $\ell_{2,1}$ norm, and $\delta > 0$ is a regularization parameter. The regularization term   $\| \nabla \cdot \|_{2,1}$ is referred to as total variation (TV) which is known to promote piecewise constant images while preserving sharp edges~\cite{rudin1992nonlinear}.  Both the $\ell_1$-norm and TV functional are not differentiable and the TV norm in particular is not well suited to gradient-based optimization due to the instability of the gradient operator. 

Alternatively, this problem can be reformulated into a constrained optimization problem of the form \begin{equation}\label{eq:l1_tv_recon}
\begin{array}{cc}
     \displaystyle{\argmin_{\x,\z_1,\z_2}} &  \|\z_1\|_1 + \delta \|\z_2\|_{2,1} \\ 
    \text{s.t. } & \mathcal{F}(\x)  - \z_1 = \d \\ 
     & \nabla \x - \z_2 = 0 
\end{array}, 
\end{equation} which can be solved utilizing multiblock ADMM.

This multiblock ADMM algorithm was applied to an image reconstruction problem where the imaging operator $\mathcal{F}$ was a sparse computed tomography (CT) imaging operator based on the Radon transform, with 20 view angles equispaced over a semicircle and 363 projections per view angle, and was implemented in Python using the SCICO package~\cite{balke2022scientific}. Measurements were then computed via the relationship  $$\d = \mathcal{F}(\x_{gt}) + \boldsymbol{\eta},$$ where $\x_{gt}$ denotes the ground truth object, a Siemens Star with 8 spokes on a $M\times M = 256\times 256$ pixel grid, and $\boldsymbol{\eta}$ represents an additive noise term corresponding to salt-and-pepper noise corrupting a quarter of the measurements.

Updating the $\x$ variable required solving an ill-conditioned linear system using conjugate gradient determined by the gradient and imaging operator. Solving this linear system comprises most of the computational burden for solving this problem. 

\begin{figure*}
    \centering
    \includegraphics[width=0.9\linewidth]{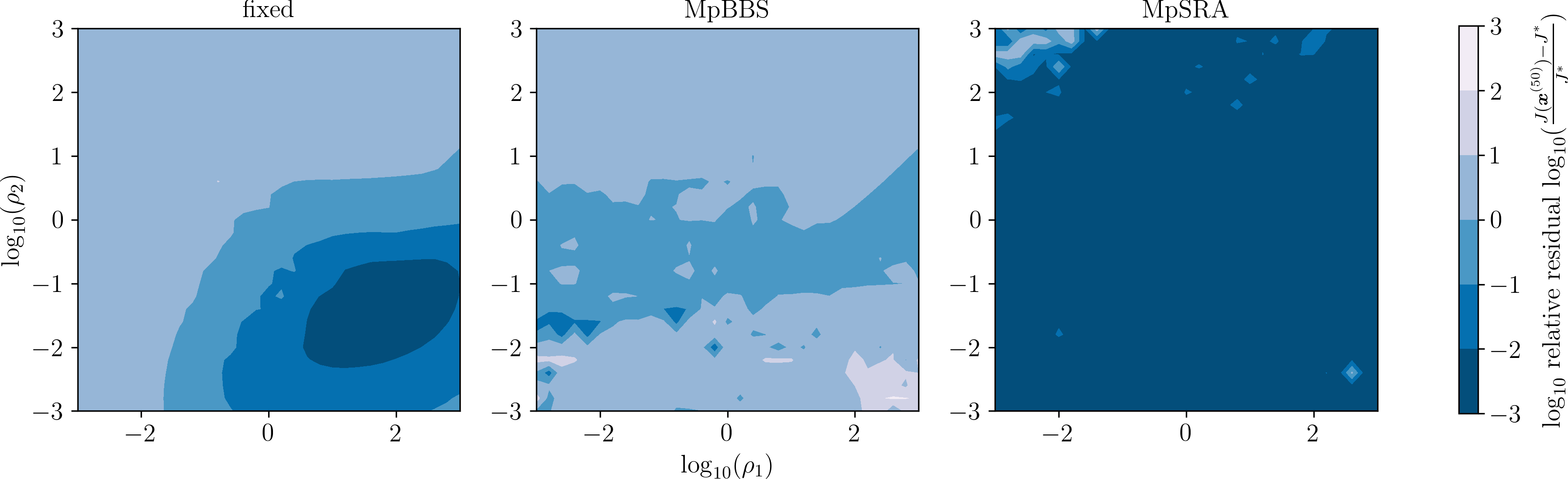}
    \caption{Relative residual of ADMM solutions for $\ell_1$ fidelity, total variation regularization sparse CT reconstruction problem utilizing fixed, multiparameter BBS, and multiparameter SRA methods after 50 iterations plotted on a surface as a function of initial $\rho_1$ and $\rho_2$. Note that the fixed method converges only in a region off of the $\rho_1 = \rho_2$ diagonal. The multiparameter BSS method demonstrates very poor performance and converges almost nowhere. The proposed multiparameter SRA method converges almost everywhere.}
    \label{fig:multiblock_l1_df_tv_reg_residual}
\end{figure*}

\begin{figure}
\centering
\includegraphics[width=0.8\columnwidth]{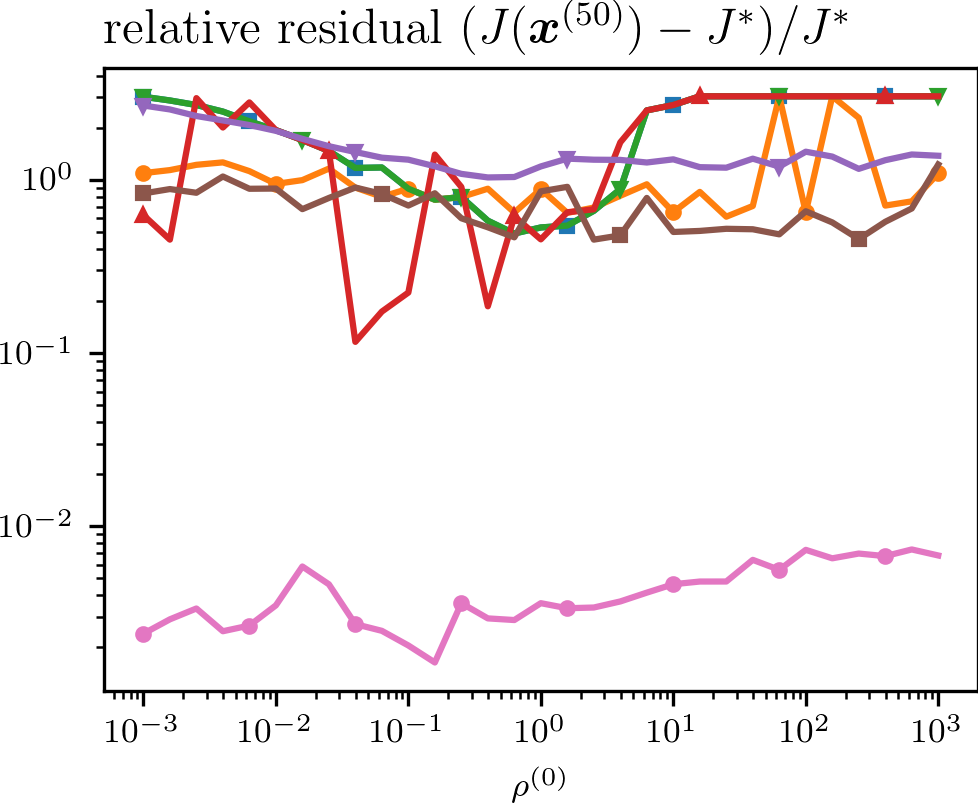}
\includegraphics[width=0.9\columnwidth]{Figures/multiblock_l1_df_total_variation_regularization_legend.pdf}
\caption{Relative residual of ADMM solutions for $\ell_1$ fidelity, total variation regularization sparse CT reconstruction problem for single-parameter and multiparameter adaptive penalty parameter rules after 50 iterations plotted as a function of initial $\rho = \rho_1 = \rho_2.$ Note that none of the single-parameter methods converge, corresponding to the optimal $\boldsymbol{\rho}$ being off-diagonal in Fig. \ref{fig:multiblock_l1_df_tv_reg_residual}. Meanwhile, the proposed multiparameter SRA method converges for all initial $\rho. $}
\label{fig:multiblock_l1_df_tv_reg_residual1}
\vspace{-5mm}
\end{figure}

\begin{figure*}
    \centering
    \includegraphics[width=0.9\linewidth]{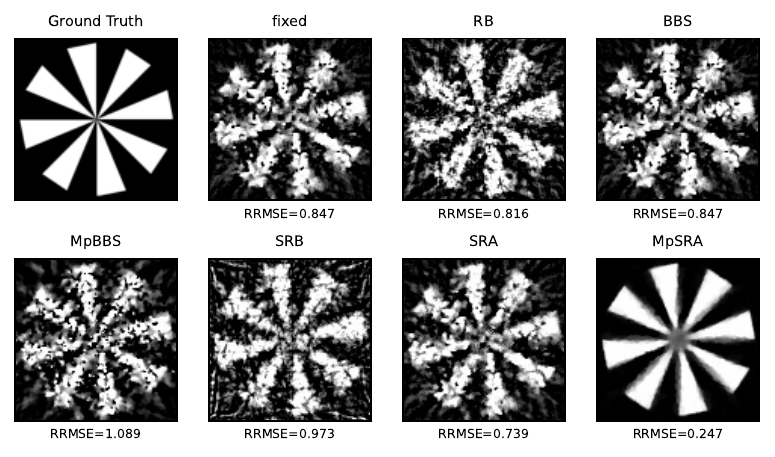}
    \caption{Example reconstructions from each penalty parameter method initialized at $\rho = \rho_1= \rho_2 = 1.$ Only the multiparameter method leads to a high-fidelity reconstruction with low relative root mean square error (RRMSE). 
    }
    \label{fig:recon_examples}
    \vspace{-5mm}
\end{figure*}

The relative residual of the multiblock algorithms for an array of starting $(\rho_1,\rho_2)$ after 50 iterations is displayed as a surface in Fig. \ref{fig:multiblock_l1_df_tv_reg_residual}. The relative residuals of the single-constraint and multiconstraint algorithms for a range of starting $\rho = \rho_1 = \rho_2$ after  50 iterations are plotted in Fig. \ref{fig:multiblock_l1_df_tv_reg_residual1}. The resulting reconstructed images from each method after 50 iterations are displayed in Fig. \ref{fig:recon_examples}. 

\subsection{Run Times}
\label{sec:run_times}

The run time of each ADMM method was recorded, compared, and presented in Table~\ref{tab:runtime}. Because the proposed method does not involve expensive computations, it does not significantly increase run time compared to the fixed method. In the image reconstruction problem, the proposed method lead to the shortest run time, most likely due to providing automatic conditioning on the linear system and requiring fewer conjugate gradient iterations.  

\begin{table*}[htbp]
    \centering
    \caption{Run time, mean $\!\pm\!$ standard deviation [s]}
    \label{tab:runtime}
    \begin{tabular}{llllllll}
    \toprule
    \toprule
    & \multicolumn{7}{c}{method}\\ \cmidrule{2-8}
        problem & fixed & RB & BBS & MpBBS & SRB & SRA & MpSRA \\\midrule
Complex Quads & 5.9e-2 $\!\pm\!$ 8.9e-3 & 6.5e-2 $\!\pm\!$ 9.8e-3 & 6.7e-2 $\!\pm\!$ 9.4e-3 & 6.6e-2 $\!\pm\!$ 9.7e-3 & 6.3e-2 $\!\pm\!$ 9.3e-3 & 6.3e-2 $\!\pm\!$ 1.0e-2 & 6.6e-2 $\!\pm\!$ 1.0e-2 \\
Scaled Quads & 1.3e-1 $\!\pm\!$ 1.4e-2 & 1.4e-1 $\!\pm\!$ 1.3e-2 & 1.4e-1 $\!\pm\!$ 1.3e-2 & 1.4e-1 $\!\pm\!$ 1.4e-2 & 1.4e-1 $\!\pm\!$ 1.3e-2 & 1.4e-1 $\!\pm\!$ 1.3e-2 & 1.4e-1 $\!\pm\!$ 1.4e-2 \\
$\ell_1$ fidelity TV reg & 1.1e{+}2 $\!\pm\!$ 6.0e+1 & 1.5e+2 $\!\pm\!$ 3.8e+1 & 1.0e+2 $\!\pm\!$ 5.0e+1 & 9.7e+1 $\!\pm\!$ 5.6e+1 & 1.5e+2 $\!\pm\!$ 1.5e+1 & 1.2e+2 $\!\pm\!$ 1.6e+1 & 7.3e+1 $\!\pm\!$ 1.0e+1 \\

\bottomrule
    \end{tabular}
    \vspace{-5mm}
\end{table*}

\subsection{Summary} Table \ref{tab:rho0} depicts the relative residual at $k=50$ with $\rho^{(0)} = \rho_1^{(0)} = \rho_2^{(0)} =1.0$ for each method across each problem.  Table \ref{tab:median} displays the median relative residual at $k=50$ for each method across each problem. The results presented in this work are consistent with empirical observations that penalty parameter selection has a large impact on convergence and optimal selection method varies between optimization problems. In the sum of quadratics problem with complex eigenvalues in the iteration matrix, the proposed method converged before the other methods.  In the scaled sum of quadratics problem, the proposed and BBS multiparameter methods did not demonstrate the best performance in the unscaled case, but they demonstrated stable performance as the scale between constraints increased while the performance of the single-parameter methods significantly decreased. This indicates that multiparameter methods that are multiscaling covariant are needed to ensure quick convergence as the scale between constraints becomes uneven. 
The results of the  $\ell_1$ data fidelity TV regularization image reconstruction problem demonstrated how each method performs on a non-quadratic, convex problem. This experiment indicated that the proposed multiparameter method demonstrated similar behavior as it did in a quadratic problem. However, the multiparameter BBS method exhibited work performance compared to the quadratic problem. In this example, multiple parameters were needed for quick and accurate reconstruction. The multiparameter BBS method did not lead to an accurate reconstruction despite a fixed optimal $\rho $ existing. The proposed MpSRA method demonstrated the best performance (by two orders of magnitude) and quickly converged regardless of initial choice of penalty parameter.

\begin{table*}[htbp]
    \centering
    \caption{Relative residual at $k=50$ with $\rho^{(0)} = \rho_1^{(0)} = \rho_2^{(0)} =1.0$}
    \label{tab:rho0}
    \begin{tabular}{llllllll}
    \toprule
    & \multicolumn{7}{c}{method}\\ \cmidrule{2-8}
        problem & fixed & RB & BBS & MpBBS & SRB & SRA & MpSRA \\\midrule
Complex Quads & 2.14e-12 & 2.14e-12 & 8.29e-9 & 1.20e-15 & 1.41e-11 & 2.41e-10 & \textbf{5.72e-16}\\
Scaled Quads($m$ = 0)  & 5.13e-4 & 1.96e-7 & 2.39e-5 & 1.94e-7 & 8.61e-9 & \textbf{1.24e-9} & 1.03e-6\\
Scaled Quads($m$ = 1)   & 2.81e-1 & 2.02e-4 & 7.37e-3 & \textbf{4.75e-8} & 1.17e-3 & 9.36e-4 & 3.90e-6 \\ 
Scaled Quads($m$ = 2) & 7.97e-1 & 3.22e-1 & 4.20e-1 & 7.10e-5 & 6.40e-1 & 5.03e-1 & \textbf{1.68e-5}\\
$\ell_1$ fidelity TV reg & 4.96e-1 & 1.09e-0 & 4.96e-1 & 4.46e-1 & 1.23e-0 & 4.92e-1 & \textbf{2.31e-3}\\
\bottomrule
    \end{tabular}
\end{table*}

\begin{table*}[htbp]
    \centering
    \caption{Median relative residual at $k=50$}
    \label{tab:median}
    \begin{tabular}{llllllll}
    \toprule
    & \multicolumn{7}{c}{method}\\ \cmidrule{2-8}
    problem & fixed & RB & BBS & MpBBS & SRB & SRA & MpSRA \\\midrule
Complex Quads & 1.32e-2 & 8.34e-11 & 8.29e-9 & \textbf{9.81e-16} & 9.71e-11 & 4.31e-10 & 1.10e-15\\
Scaled Quads($m$ = 0)  & 1.40e-1 & 1.36e-7 & 1.52e-5 & 1.24e-6 & 3.66e-8 & \textbf{3.00e-9} & 3.97e-6\\
Scaled Quads($m$ = 1)  & 3.55e-1 & 3.79e-4 & 5.12e-3 & \textbf{2.78e-6} & 1.17e-3 & 1.01e-3 & 6.76e-6\\ 
Scaled Quads($m$ = 2)  &  7.97e-1 & 2.85e-1 & 4.09e-1 & \textbf{1.25e-5} & 6.40e-1 & 4.80e-1 & 1.39e-5\\
$\ell_1$ fidelity TV reg & 2.50e+00 & 8.85e-1 & 2.50e+00 & 1.97e+00 & 1.25e+00 & 6.69e-1 & \textbf{3.86e-3}\\
\bottomrule
    \end{tabular}
\end{table*}

\section{Conclusions}\label{sec:conclusion}

This work proposes an adaptive multiparameter selection method for ADMM applied to multiconstraint and multiblock optimization problems. This method was developed via a theoretical framework that analyzes ADMM as an affine fixed-point iteration problem and attempts to minimize the spectral radius of the iteration matrix involved. The proposed multiparameter method is referred to as the multiparameter spectral radius approximation (MpSRA) method and is an extension of the single-parameter SRA method ~\cite{mccann_robust_2024} and is derived by extending and correcting the analysis of \cite{mccann_robust_2024} for spectral approximation and optimization with multiple parameters. The MpSRA method is intended to be simple to understand and implement while providing robust performance with respect to initial parameters and preventing further complexity associated with multiple parameters. 

The efficacy of the proposed method was demonstrated and compared to several single-parameter ADMM approaches and another multiparameter method, the multiparameter  Barzilai-Borwein spectral (BBS) method, in 
three numerical experiments. In each of these experiments, each adaptive penalty parameter method for ADMM was applied to an optimization problem. The first optimization problem was a constrained sum of quadratics optimization problem which resulted in an iteration matrix with complex eigenvalues. This experiment demonstrated that the proposed method assumptions around complex eigenvalues held and that the proposed method converges quicker than the other methods regardless of initial penalty parameter. The second optimization problem was a sum of quadratics with multiple constraints, and the associated experiment was repeated for constraints at three different scales. This experiment demonstrates that a multiparameter method may not lead to the quickest convergence when constraints are scaled equally, but, as the scales between constraints grow, a multiparameter method is necessary for quick convergence. 
The third optimization problem was a multiblock optimization formulation of image reconstruction for sparse computed tomography using an $\ell_1$ data fidelity and TV regularization. This problem was an example of an ADMM problem in which the quickest convergence could not be achieved with a single-parameter and a multiparameter method was needed. The proposed method was the only method that achieved satisfactory convergence within 50 iterations and presented little dependence on initial penalty parameter. These experiments highlight the empirical performance of the proposed method and that is competitive or superior to state-of-the-art methods.

\appendices
\section{Equivalence of Multiconstraint ADMM and Standard ADMM via Diagonal Preconditioning}\label{app:multi_equivalence}

Consider the constrained optimization problem equivalent to the one in \eqref{eq:multiconstraint_admm_problem} with scaled constraints \begin{equation} \label{eq:multiconstraint_admm_problem_scaled}
\begin{array}{ccc}
\displaystyle{\argmin_{\x, \z_{1},\hdots, \z_J}}  & f(\x) + g(\z)&  \\ 
\text{s.t.} &  \beta_j\A_j\x + \beta_j\B_j\z = \beta_j\c_j & j = 1,\hdots, J ,
\end{array}
\end{equation} for scalar parameters $\{\beta_j\}_{j=1}^J.$ Letting $\boldsymbol{\beta} = (\beta_1,\hdots,\beta_J)^T$ and $\D$ be the diagonal operator in \eqref{eq:diagonal_matrix_operator}, then \eqref{eq:multiconstraint_admm_problem_scaled} can be expressed in a form with a single constraint \begin{equation} \label{eq:multiconstraint_admm_problem_scaled_single}
\begin{array}{cc}
\displaystyle{\argmin_{\x, \z_{1},\hdots, \z_J}}  & f(\x) + g(\z)  \\ 
\text{s.t.} &  \D_{\boldsymbol{\beta}}\A\x +  \D_{\boldsymbol{\beta}}\B\z =  \D_{\boldsymbol{\beta}}\c .
\end{array}
\end{equation} 

The standard implementation of ADMM can then be applied to this single-constraint problem with a scalar penalty parameter $\rho_0 >0$ \begin{equation}\label{eq:Scaled_Standard_ADMM_x}
  \begin{array}{ll}
       \x^{(k\!+\!1)} & = \displaystyle{\argmin_{\x}} \  f(\x) \\ 
       & \!+\! \frac{\rho_0}{2} \left\| \D_{\boldsymbol{\beta}}\A\x \!+\! \D_{\boldsymbol{\beta}} \B\z^{(k)} \!-\! \D_{\boldsymbol{\beta}}\c + \frac{\tilde{\y}^{(k)}}{\rho_0} \right\|^2 
  \end{array}
  \end{equation} \begin{equation}\label{eq:Scaled_Standard_ADMM_z}
  \begin{array}{ll}
       \z^{(k\!+\!1)} & = \displaystyle{\argmin_{\z}} \  g(\z)\\ 
       &  \!+\! \frac{\rho_0}{2} \left\| \D_{\boldsymbol{\beta}}\A\x^{(k+1)} \!+\! \D_{\boldsymbol{\beta}}\B\z \!-\! \D_{\boldsymbol{\beta}}\c + \frac{\tilde{\y}^{(k)}}{\rho_0} \right\|^2 
  \end{array}
  \end{equation} \begin{equation}\label{eq:Scaled_Standard_ADMM_tilde_y}
  \begin{array}{ll}
    \tilde{\y}^{(k\!+\!1)} & = \tilde{\y}^{(k)} \\
    & + \rho_0 \big( \D_{\boldsymbol{\beta}} \A\x^{(k\!+\!1)} + \D_{\boldsymbol{\beta}}\B\z^{(k\!+\!1)} - \D_{\boldsymbol{\beta}}\c \big) .
  \end{array}      
  \end{equation}
Letting $\y^{(k)} = \D_{\boldsymbol{\beta}}\tilde{y}^{(k)}$ and splitting the constraints reformulates Eqs. \eqref{eq:Scaled_Standard_ADMM_x}-\eqref{eq:Scaled_Standard_ADMM_tilde_y} to \begin{equation}\label{eq:Unscaled_Standard_ADMM_x}
        \begin{array}{ll}
             \x^{(k\!+\!1)} & = \displaystyle{\argmin_{\x}} \  f(\x)\\ 
             &  \!+\! \displaystyle{\sum_{j=1}^J \frac{\rho_0\beta_j^2}{2} \left\| \A_j\x \!+\!  \B_j\z^{(k)} \!-\! \c_j + \frac{{\y}^{(k)}_j}{\rho_0\beta_j^2} \right\|^2 }
        \end{array}
    \end{equation}   \begin{equation}
    \label{eq:Unscaled_Standard_ADMM_z}
    \begin{array}{ll}
          \z^{(k\!+\!1)} & = \displaystyle{\argmin_{\z}} \  g(\z)\\ 
         &  \!+\! \displaystyle{\sum_{j=1}^J\frac{\rho_0\beta_j^2}{2} \left\| \A_j\x^{(k+1)} \!+\! \B_j\z \!-\! \c_j + \frac{{\y_j}^{(k)}}{\rho_0\beta_j^2} \right\|^2} 
    \end{array}
    \end{equation} \begin{equation}\label{eq:Unscaled_Standard_ADMM_yj}
    \begin{array}{ll}
    {\y}^{(k\!+\!1)}_j & = {\y_j}^{(k)} + \rho_0\beta_j^2 \big(  \A_j\x^{(k\!+\!1)} + \B_j\z^{(k\!+\!1)} - \c_j \big) .
    \end{array}        
    \end{equation}
Letting $\rho_j = \rho_0 \beta_j^2$ recovers the multiparameter ADMM implementation in Eqs. \eqref{eq:Multiconstraint_ADMM_x}-\eqref{eq:Multiconstraint_ADMM_yj}. 

Thus the multiparameter ADMM for multiconstraint problems is equivalent to the standard implementation of ADMM with scaled constraints. This multiparameter method then inherits the convergence properties of standard ADMM. Furthermore, this formulation implies that finding the optimal set of penalty parameters $\boldsymbol{\rho} = (\rho_1,\hdots,\rho_J)^T$  in Eqs. \eqref{eq:Multiconstraint_ADMM_x}-\eqref{eq:Multiconstraint_ADMM_yj} is equivalent to finding a single optimal penalty parameter $\rho_0$ and diagonal scaling or conditioning between constraints \cite{giselsson-2014-diagonal} $\boldsymbol{\beta} = (\beta_1,\hdots,\beta_J)^T$ in Eqs. \eqref{eq:Scaled_Standard_ADMM_x}-\eqref{eq:Scaled_Standard_ADMM_tilde_y}. Note that this work focuses on scaling between constraints, but these same ideas could be expanded to scaling constraints with a matrix and introducing a formulation of ADMM based on norms with positive definite matrices.

\bibliographystyle{IEEEtranD}
\bibliography{refs_FINAL, refs_other}

\end{document}